\documentclass[12pt]{iopart}

\usepackage{graphicx}
\usepackage{iopams}
\usepackage{subfigure}
\usepackage{arydshln}

\newcommand{\bra}[1] {\langle #1 |}
\newcommand{\ket}[1] {| #1 \rangle}
\newcommand{\braket}[2] {\langle #1 | #2 \rangle}

\newcommand{\cg}[6] {\left(\begin{array}{cc|c} #1 & #3 & #5  \\ #2 & #4 & #6 \end{array}\right)}
\newcommand{\rdm}[3]{\bra{#1}| #2 | \ket{#3}}
\newcommand{\me}[3]{\bra{#1} #2 \ket{#3}}

\begin{document}

\title[The electronic solution of the NV$^-$ centre]{The negatively charged nitrogen-vacancy centre in diamond: the electronic solution}

\author{M W Doherty$^1$, N B Manson$^2$, P Delaney$^3$ and L C L Hollenberg$^1$}

\address{$^1$ School of Physics, University of Melbourne, Victoria 3010, Australia.}
\address{$^2$ Laser Physics Centre, Research School of Physical Sciences and Engineering, Australian National University, Australian Capital Territory 0200, Australia.}
\address{$^3$ School of Mathematics and Physics, Queen's University Belfast, Northern Ireland BT7 1NN, United Kingdom.}
\ead{marcuswd@unimelb.edu.au}

\begin{abstract}
The negatively charged nitrogen-vacancy centre is a unique defect in diamond that possesses properties highly suited to many applications, including quantum information processing, quantum metrology, and biolabelling. Although the unique properties of the centre have been extensively documented and utilised, a detailed understanding of the physics of the centre has not yet been achieved. Indeed there persists a number of points of contention regarding the electronic structure of the centre, such as the ordering of the dark intermediate singlet states. Without a detailed model of the centre's electronic structure, the understanding of the system's unique dynamical properties can not effectively progress. In this work, the molecular model of the defect centre is fully developed to provide a self consistent model of the complete electronic structure of the centre. The application of the model to describe the effects of electric, magnetic and strain interactions, as well as the variation of the centre's fine structure with temperature, provides an invaluable tool to those studying the centre and a means to design future empirical and \textit{ab initio} studies of this important defect.
\end{abstract}
\submitto{\NJP}
\maketitle

\section{Introduction}

The negatively charged nitrogen-vacancy (NV$^-$) centre  in diamond is a promising system for many quantum information processing \cite{qip}, quantum metrology, and biolabelling applications \cite{bio}. These applications include secure quantum key distribution \cite{qkd}, quantum computing \cite{comp}, Q-switching in solid state photonic cavities \cite{qswitch}, magnetometry \cite{mag}, electric field sensing \cite{efield} and decoherence based imaging \cite{decoherence}. The significant interest in the centre is primarily due to its well documented capabilities of single-photon generation \cite{singlephoton}, long-lived coherence \cite{coherence}, spin coupling \cite{coupling} and optical spin polarization and readout \cite{readout}. The observed properties of the centre include a strong optical zero phonon line (ZPL) at 1.945 eV \cite{dupreez}, an infrared ZPL at 1.190 eV \cite{infrared}, a paramagnetic ground state triplet \cite{ground}, and a strain \cite{excitedstatestrain}, Zeeman \cite{excitedstatezeeman} and Stark \cite{tamarat} affected excited state triplet. Recent experimental studies have also provided new information regarding the centre's excited state fine structure, its temperature dependence \cite{averaging} and the presence of dynamic Jahn-Teller effects \cite{jahnteller}.

\begin{figure}[hbtp]
\begin{center}
\includegraphics[width=0.66\columnwidth] {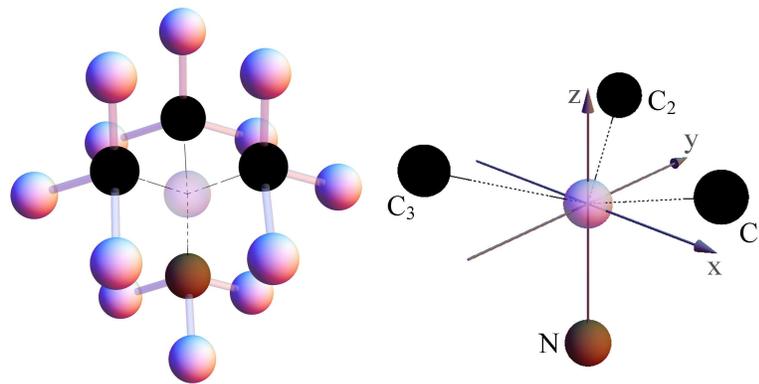}
\caption{Schematics of the nitrogen-vacancy centre and lattice depicting: the vacancy (transparent); the nearest-neighbour carbon atoms to the vacancy (black); the substitutional nitrogen atom (brown); and, the next-to-nearest carbon neighbours to the vacancy (white). The adopted coordinate system and carbon labels are depicted in the right schematic.}
\label{fig:center}
\end{center}
\end{figure}

The NV$^-$ centre is a point defect of $C_{3v}$ symmetry in diamond consisting of a substitutional nitrogen atom adjacent to a carbon vacancy (refer to \fref{fig:center}). The observable properties of the centre are consistent with a six electron model \cite{loubser}, where the electrons are postulated to consist of the five unpaired electrons of the nearest neighbour nitrogen and carbon atoms to the vacancy and an additional electron trapped at the centre. The even number of electrons yields an integer spin system and the use of spin resonance techniques \cite{xing} has confirmed that the electronic states are highly localized to the vacancy and its nearest neighbours and that the ground state is an $^3A_2$ triplet \cite{ground}. The high degree of localization supports the application of a molecular model of the electronic system of the centre, in which the centre's electronic states are described by configurations of molecular orbitals (MOs) initially constructed from linear combinations of the dangling $sp^3$ orbitals of the nearest neighbour carbon and nitrogen atoms using group theoretical arguments.

The previous applications of the molecular model \cite{loubser,lenef,manson} have successfully described the zero field splitting of the ground triplet state due to spin-spin interaction, the $^3E$ excited triplet state and its fine structure induced by both spin-orbit and spin-spin interactions, and many aspects of the interactions of the triplet states with electric, magnetic and strain perturbations. However, being a semi-empirical model, unless the molecular model is fully developed in order to reduce the model's parameters to the minimal set, the model has limited ability to make definitive predictions on aspects of the electronic structure which can not be directly observed and the design of a systematic method to measure the large set of parameters becomes ambiguous. As a result of the previous partially developed molecular models, there has been a number of continual points of contention regarding the electronic structure of the centre. In particular, the contention surrounding the energetic ordering and positioning of the dark $^1E$ and $^1A_1$ (and possibly $^1E'$) singlet states that are thought to exist between the ground and excited triplet states.

The locations of the intermediate singlet states are critical to developing an understanding of the process of optical spin polarization \cite{readout}, which is the principle property of the NV$^-$ centre that underpins the majority of its important applications. In the process of optical spin polarization, the population which has been optically excited from the ground to the excited triplet state is believed to decay non-radiatively from the excited triplet via the intermediate singlets to the ground triplet state in such a way that the $m_s = \pm1$ sub-levels of the excited triplet state are preferentially depopulated and the $m_s = 0 $ sub-level of the ground triplet is preferentially populated. Consequently, after a short period of optical excitation the centre becomes spin-polarized into $m_s = 0$ population. This paper does not aim to describe the spin polarization mechanism, but instead provide the detailed model of the coupling of the intermediate singlet states and the triplet states due to spin-orbit interactions, which will form the foundations for future studies of the spin polarization mechanism.

There are a number of other properties of the NV$^-$ centre that require a fully developed model in order to be satisfactorily explained. These include the Stark effect in the ground state triplet \cite{efield, vanoort}, the small anisotropy of the effective electronic g-factor of the ground state triplet \cite{gfactor}, the strain splitting of the infrared transition between the intermediate singlet states \cite{infrared}, and the presence of the Jahn-Teller effect in the $^1E$ and $^3E$ \cite{jahnteller}. Each of these properties requires the detailed treatment of electronic Coulomb repulsion, spin-orbit and spin-spin interactions which act to couple the electronic states of the centre and allow these properties to exist. The coupling of electronic states implies that the calculation of the effects of electric, magnetic and strain perturbations must be conducted in the complete basis of the centre's electronic states.

In this article, the molecular model of the NV$^-$ centre will be fully developed to provide an electronic solution that is experimentally testable and offers explanations for many of the remaining questions regarding the centre. The model will be based upon previous applications of the molecular model and utilize invaluable \textit{ab initio} and empirical results to draw conclusions and identify parameters which are known. The electronic Coulomb, spin-orbit and spin-spin interactions will be treated to determine the energies and couplings of the electronic states. Spin-orbit and spin-spin interactions will be treated using perturbation theory in order to produce simple energy and coupling coefficient expressions in terms of the minimal set of model parameters. Each of the parameters are provided as one- and two-electron matrix elements of the electronic interactions, allowing unambiguous evaluation by future \textit{ab initio} studies. Electric, magnetic and strain interactions are also treated in order to allow future experiments to be designed to measure the remaining unknown parameters. The treatment of the interactions will also provide the foundations to develop an understanding of spin polarization, the Jahn-Teller effect and the temperature dependence of the centre's properties.

\section{The orbital structure}

Adopting an adiabatic approximation and considering the nuclei of the crystal to be fixed at their equilibrium coordinates $\vec{R}_0$ corresponding to the ground electronic state, the electronic Hamiltonian of the NV$^-$ centre can be defined as
\begin{eqnarray}
\hat{H}_{NV} = \sum_i\hat{T}_i+\hat{V}_{Ne}(\vec{r}_i,\vec{R}_0)+\hat{V}_{so}(\mathbf{x}_i,\vec{R}_0)+\sum_{i>j}\hat{V}_{ee}(\mathbf{x}_i,\mathbf{x}_j)
+\hat{V}_{ss}(\mathbf{x}_i,\mathbf{x}_j) \nonumber \\
\label{eq:NVhamiltonian}
\end{eqnarray}
where $\mathbf{x}_i = (\vec{r}_i,\vec{s}_i)$ denotes the collective spatial and spin coordinates of the $i^{th}$ electron of the centre, $\hat{T}_i$ is the kinetic energy of the $i^{th}$ electron, $\hat{V}_{Ne}$ is the effective Coulomb potential of the interaction of the nuclei and lattice electrons with the electrons of the centre, $\hat{V}_{so}$ is the electronic spin-orbit potential, $\hat{V}_{ee}$ is the Coulomb repulsion potential of the electrons of the centre, and $\hat{V}_{ss}$ is the electronic spin-spin potential. Note that nuclear hyperfine interactions have been ignored. As for most molecular and solid state systems, the first step in solving $\hat{H}_{NV}$ is to obtain the solutions of the one-electron Coulomb problem,
\begin{eqnarray}
\hat{h} = \hat{T}+\hat{V}_{Ne}(\vec{r},\vec{R}_0)
\label{eq:oneelectronhamiltonian}
\end{eqnarray}
which will be the MOs of the centre. Using the MOs, a basis of many-electron configuration states that are solutions of $\sum_i\hat{h}_i$ can be defined and the remaining one- and two-electron components of $\hat{H}_{NV}$ can be treated in this basis.

At this stage, the $C_{3V}$ symmetry of the ground nuclear equilibrium coordinates can be employed to construct the MOs of the defect. Using the basis $\{n, c_1, c_2, c_3\}$ (refer to \fref{fig:center} for labels) of tetrahedrally coordinated $sp^3$ atomic orbitals of the nearest neighbour carbon and nitrogen atoms to the vacancy, the MOs can be constructed as linear combinations of the atomic orbitals (LCAOs) with definite orbital symmetry. This procedure has been conducted by a number of authors and an example of the resulting set of MOs \cite{prl} is
\begin{eqnarray}
a_1(N) = n, \ \ a_1(C) = \frac{1}{\sqrt{3}\sqrt{1+2S_{cc}-3S_{nc}^2}}(c_1+c_2+c_3-3 S_{nc} n), \nonumber\\
e_x = \frac{1}{\sqrt{3}\sqrt{2-2S_{cc}}}(2c_1-c_2-c_3), \ \ e_y = \frac{1}{\sqrt{2-2S_{cc}}}(c_2-c_3)\label{eq:MOs}
\end{eqnarray}
where $S_{nc} = \braket{n}{c_1}$ and $S_{cc} = \braket{c_1}{c_2}$ are orbital overlap integrals.

Clearly the LCAO method is a highly approximate method of constructing the MOs as it uses a restricted basis set and does not consider the interactions between the MOs of the defect centre and the electron orbitals of the remainder of the crystal. Therefore, the key objective of the LCAO method is not to produce an accurate description of the MOs, but to produce the correct number of MOs of a particular symmetry type and to estimate their energy ordering. The results of \textit{ab initio} studies \cite{ab initio, delaney, gali3} can be used to confirm the number of MOs of each symmetry type and their energy ordering. The majority of \textit{ab initio} studies agree that there exists three MOs $\{a_1, e_x, e_y\}$ within the bandgap of diamond and that these resemble the highly localized MOs of \eref{eq:MOs}, with additional contributions from atomic orbitals at the next-to-nearest neighbours and beyond. Furthermore, the studies show that the $a_1(N)$ and $a_1(C)$ MOs have mixed to form $a_1$ and $a_1'$ such that $a_1'$ has been forced downwards in energy into the diamond valence band and $a_1$ has significant contributions from both the nitrogen and carbon atomic orbitals.

\begin{figure}[hbtp]
\begin{center}
\includegraphics[width=0.95\columnwidth] {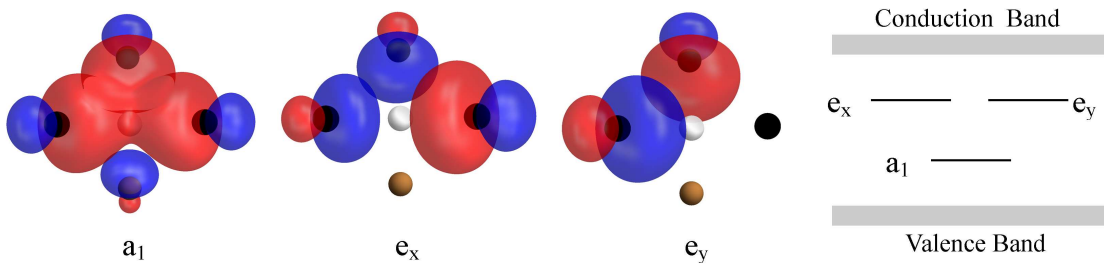}
\caption{Schematics of the three NV centre molecular orbitals (MOs) responsible for the centre's observable properties in the vicinity of the vacancy and their energy ordering. Red and blue components represent positive and negative contributions to the MO respectively.}
\label{fig:MOs}
\end{center}
\end{figure}

Using the six electron model of the NV$^-$ centre, the $a_1'$ MO will be completely filled by two electrons in the ground $a_1'^2a_1^2e^2$ and first excited $a_1'^2a_1^1e^3$ MO configurations. There are several second excited MO configurations: $a_1'^2e^4$, $a_1'a_1^2e^3$ and $a_1'a_1e^4$. Due to the estimated proximity of the $a_1$ MO to the valence band \cite{prl,gali3}, the first two second excited MO configurations $a_1'^2e^4$ and  $a_1'a_1^2e^3$ could be close in energy. The states of these second excited MO configurations could mix with the states of the ground and first excited states and affect their energies, but are not expected to play a significant role in the centre's properties themselves. Consequently, only the three MOs within the band gap are expected to contribute to the observable properties of the centre and only  the ground and first excited MO configurations will be treated in detail in this work. Schematics of the three MOs in the region of the vacancy and their energy ordering is depicted in \fref{fig:MOs}.

The configuration states of the ground and first excited MO configurations of the centre are constructed by first defining linear combinations of products of four MOs that transform as a particular row of an irreducible representation of the $C_{3v}$ group. Using the irreducible representations and Clebsch-Gordon coefficients defined in \cite{lenef}, examples of the symmeterized linear combinations of products of two MOs are
\begin{eqnarray}
\phi_{A_1}(a_1a_1) = a_1a_1, \ \ \phi_{E,x}(a_1e) = a_1e_x, \ \ \phi_{E,y}(a_1e) = a_1e_y, \nonumber \\
\phi_{A_1}(ee) = \frac{1}{\sqrt{2}}(e_xe_x+e_ye_y), \ \ \phi_{A_2}(ee) = \frac{1}{\sqrt{2}}(e_xe_y-e_ye_x), \nonumber \\
\phi_{E,x}(ee) = \frac{1}{\sqrt{2}}(e_xe_x-e_ye_y), \ \ \phi_{E,y}(ee) = \frac{-1}{\sqrt{2}}(e_xe_y+e_ye_x)
\label{eq:twoMOproducts}
\end{eqnarray}
The symmeterized combinations of products of four MOs can be constructed by repeating the process used to construct the above products of two MOs. Once the symmeterized products of occupied MOs corresponding to each configuration state are constructed, the configuration states are formed by performing a direct product with an associated spin state and transforming the result into a linear combination of Slater determinants. The resulting configuration states $\Phi_{j,k;S,m_s}^c$ have definite orbital symmetry ($j$ denoting irreducible representation and $k$ denoting row of the irreducible representation), total spin $S$, and spin projection $m_s$, and are explicitly contained in \tref{tab:states}. Note that the construction of the configuration states in this way is completely analogous to LS coupling in atomic structure, where the atomic states are constructed to have definite orbital $(L,m_l)$ and spin $(S,m_s)$ quantum numbers prior to the introduction of spin-orbit interaction. The configuration states may also be expressed in terms of holes rather than electrons and these expressions are contained in \tref{tab:holes}. However, the hole formulism will not be used in the remainder of this article.

\begin{table}
\caption{\label{tab:states}The configuration and spin-orbit states of the NV$^-$ centre expressed in terms of Slater determinants of the molecular orbitals. Second quantization notation has been adopted to denote the occupation of the molecular orbitals in each Slater determinant in the order $\ket{a_1\bar{a}_1e_x\bar{e}_xe_y\bar{e}_y}$, where an overbar denotes spin-down.}
\begin{indented}
\item[]\begin{tabular}{lllcllcl}
\br
 & & $\Phi_{j,k;S,m_s}^c$ & & & $\Phi_{n,j,k}^{so}$ & & \\
\mr
 $a_1^2e^2$ & $^3A_2$ & $\Phi_{A_2;1,0}^c$ & = & $\frac{1}{\sqrt{2}}(\ket{111001}+\ket{110110})$ & $\Phi_{1,A_1}^{so}$ & = & $\Phi_{A_2;1,0}^c$ \\
  & & $\Phi_{A_2;1,1}^c$ & = & $\ket{111010}$ & $\Phi_{2,E,x}^{so}$ & = & $\frac{-1}{\sqrt{2}}(-\Phi_{A_2;1,1}^c+\Phi_{A_2;1,-1}^c)$ \\
  & & $\Phi_{A_2;1,-1}^c$ & = & $\ket{110101}$ & $\Phi_{2,E,y}^{so}$ & = & $\frac{-i}{\sqrt{2}}(\Phi_{A_2;1,1}^c+\Phi_{A_2;1,-1}^c)$ \\
 & $^1E$ & $\Phi_{E,x;0,0}^c$ & = & $\frac{1}{\sqrt{2}}(\ket{111100}-\ket{110011})$ & $\Phi_{3,E,x}^{so}$ & = & $\Phi_{E,x;0,0}^{c}$ \\
  & & $\Phi_{E,y;0,0}^c$ & = & $\frac{1}{\sqrt{2}}(\ket{110110}-\ket{111001})$ & $\Phi_{3,E,y}^{so}$ & = & $\Phi_{E,y;0,0}^{c}$ \\
  & $^1A_1$ & $\Phi_{A_1;0,0}^c$ & = & $\frac{1}{\sqrt{2}}(\ket{111100}+\ket{110011})$ & $\Phi_{4,A_1}^{so}$ & = & $\Phi_{A_1;0,0}^c$ \\
 $a_1e^3$ & $^3E$ & $\Phi_{E,x;1,0}^c$ & = & $\frac{1}{\sqrt{2}}(\ket{100111}+\ket{011011})$ & $\Phi_{5,E,x}^{so}$ & = & $\frac{1}{2}\left[-i(\Phi_{E,x;1,1}^c+\Phi_{E,x;1,-1}^c))\right.$ \\
 & & & & & & & $\left.-(-\Phi_{E,y;1,1}^c+\Phi_{E,y;1,-1}^c)\right]$ \\
 & & $\Phi_{E,y;1,0}^c$ & = & $\frac{1}{\sqrt{2}}(\ket{101101}+\ket{011110})$ & $\Phi_{5,E,y}^{so}$ & = & $\frac{1}{2}\left[-(-\Phi_{E,x;1,1}^c+\Phi_{E,x;1,-1}^c)\right.$ \\
 & & & & & & & $\left.+i(\Phi_{E,y;1,1}^c+\Phi_{E,y;1,-1}^c)\right]$ \\
 & & $\Phi_{E,x;1,1}^c$ & = & $\ket{101011}$ & $\Phi_{6,E,x}^{so}$ & = & $-\Phi_{E,y;1,0}^c$ \\
 & & $\Phi_{E,y;1,1}^c$ & = & $\ket{101110}$ & $\Phi_{6,E,y}^{so}$ & = & $\Phi_{E,x;1,0}^c$ \\
 & & $\Phi_{E,x;1,-1}^c$ & = & $\ket{010111}$ & $\Phi_{7,A_2}^{so}$ & = & $\frac{1}{2}\left[(-\Phi_{E,x;1,1}^c+\Phi_{E,x;1,-1}^c)\right.$ \\
 & & & & & & & $\left.+i(\Phi_{E,y;1,1}^c+\Phi_{E,y;1,-1}^c)\right]$ \\
 & & $\Phi_{E,y;1,-1}^c$ & = & $\ket{011101}$ & $\Phi_{8,A_1}^{so}$ & = & $\frac{1}{2}\left[-i(\Phi_{E,x;1,1}^c+\Phi_{E,x;1,-1}^c))\right.$ \\
 & & & & & & & $\left.+(-\Phi_{E,y;1,1}^c+\Phi_{E,y;1,-1}^c)\right]$ \\
 & $^1E'$ & $\Phi_{E',x;0,0}^c$ & = & $\frac{1}{\sqrt{2}}(\ket{100111}-\ket{011011})$ & $\Phi_{9,E,x}^{so}$ & = & $\Phi_{E',x;0,0}^{c}$ \\
 & & $\Phi_{E',y;0,0}^c$ & = & $\frac{1}{\sqrt{2}}(\ket{101101}-\ket{011110})$ & $\Phi_{9,E,y}^{so}$ & = & $\Phi_{E',y;0,0}^{c}$ \\
\br
\end{tabular}
\end{indented}
\end{table}

\begin{figure}[hbtp]
\begin{center}
\includegraphics[width=0.65\columnwidth] {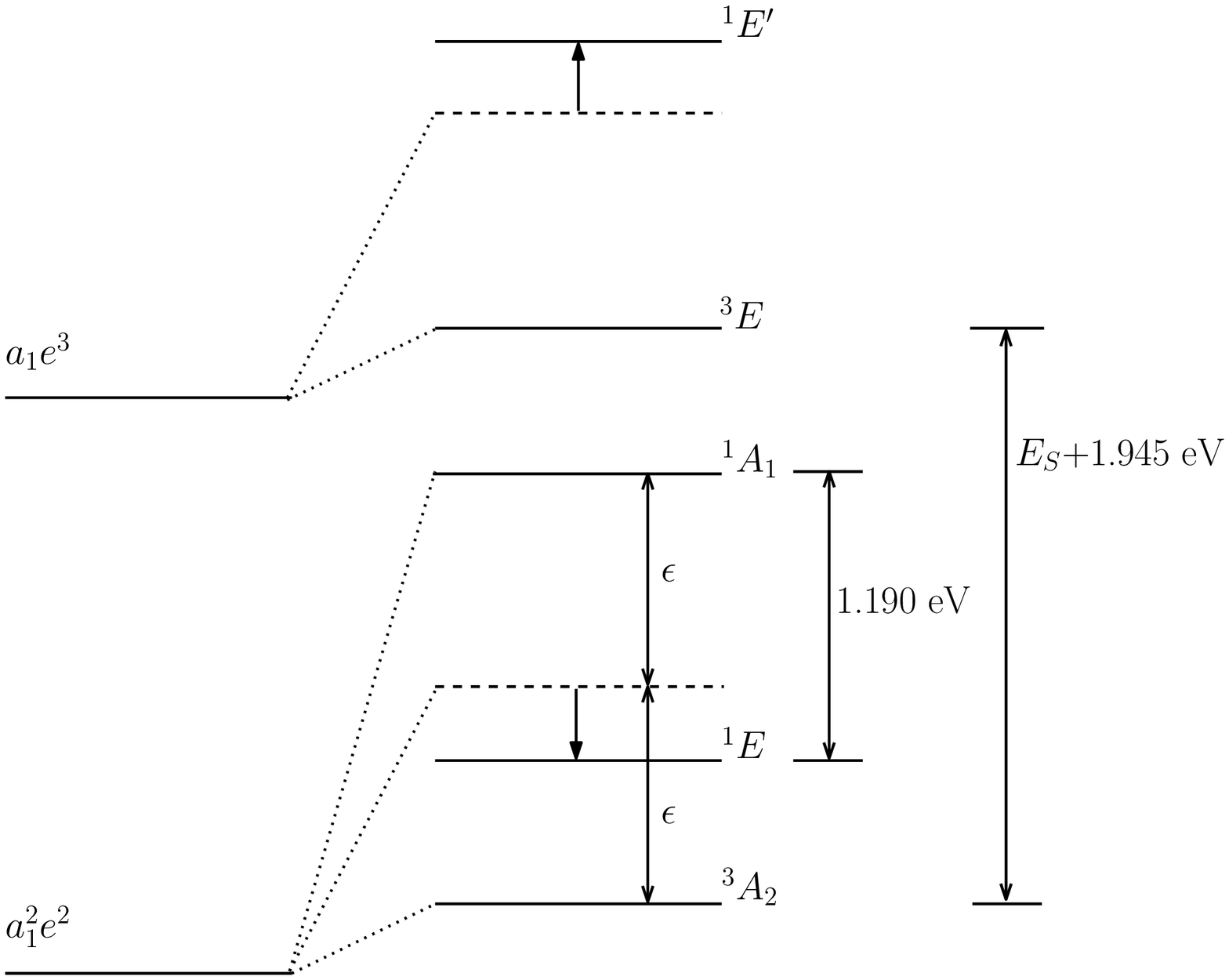}
\caption{Energy level diagram of the orbital structure of the NV$^-$ centre. The molecular orbital (MO) configuration energies are depicted on the left hand side and the splittings into singlet and triplets due to the introduction of electronic Coulomb repulsion is depicted on the right hand side. The dashed lines indicate the locations of the $E$ singlet states prior to their Coulomb coupling. $\epsilon = 2\me{e_xe_x}{\hat{V}_{ee}}{e_ye_y}$ is the difference in the Coulomb energies between the ground MO configuration states prior to the coupling of the $E$ singlets. $E_S \approx 0.235$ eV is the Stokes shift of the optical ZPL.}
\label{fig:orbitalstructure}
\end{center}
\end{figure}

The configuration states are solutions of $\sum_i\hat{h}_i$, with each of the states of a MO configuration having the same eigenenergy as depicted on the left hand side of \fref{fig:orbitalstructure}. Employing the Wigner-Eckart theorem \cite{cornwell},
\begin{eqnarray}
\me{\phi_{f,g}}{\hat{O}_{p,q}}{\phi_{j,k}} = \cg{j}{k}{p}{q}{f}{g}^\ast\rdm{\phi_f}{\hat{O}_p}{\phi_j}
\end{eqnarray}
where $\hat{O}$ is a tensor operator, $(g,q,k)$ denote the rows of the irreducible representations $(f,p,j)$ of the $C_{3v}$ group respectively and $\rdm{}{}{}$ is the reduced density matrix element, the eigenenergies of each MO configuration can be expressed in terms of reduced density matrix elements involving the MOs
\begin{eqnarray}
a_1^2e^2: \ \ 2\rdm{a_1}{\hat{h}}{a_1}+2\rdm{e}{\hat{h}}{e}, \ \ a_1e^3: \ \ \rdm{a_1}{\hat{h}}{a_1}+3\rdm{e}{\hat{h}}{e}
\label{eq:oneelectronenergies}
\end{eqnarray}
The introduction of  the Coulomb repulsion potential $\sum_{i>j}\hat{V}_{ee}(\mathbf{x}_i,\mathbf{x}_j)$ splits the MO configurations into distinct triplet and singlet energy levels. The diagonal matrix elements of the ground MO configuration triplet and singlets are
\begin{eqnarray}
^3A_2: \ \ C_0+ \rdm{\phi_{A_2}(ee)}{\hat{V}_{ee}}{\phi_{A_2}(ee)} \nonumber \\
^1E: \ \ C_0+ \rdm{\phi_{E}(ee)}{\hat{V}_{ee}}{\phi_{E}(ee)} \nonumber \\
^1A_1: \ \ C_0+ \rdm{\phi_{A_1}(ee)}{\hat{V}_{ee}}{\phi_{A_1}(ee)} \nonumber
\end{eqnarray}
where
\begin{eqnarray}C_0 = \rdm{\phi_{A_1}(a_1a_1)}{\hat{V}_{ee}}{\phi_{A_1}(ee)}+4\rdm{\phi_{E}(a_1e)}{\hat{V}_{ee}}{\phi_{E}(a_1e)}-2\rdm{\phi_{E}(a_1e)}{\hat{V}_{ee}}{\phi_{E}(ea_1)}\nonumber
\end{eqnarray}
Hund's rules indicate that the $^3A_2$ triplet is the ground electronic state, which implies that $\rdm{\phi_{A_2}(ee)}{\hat{V}_{ee}}{\phi_{A_2}(ee)} < $ $\rdm{\phi_{E}(ee)}{\hat{V}_{ee}}{\phi_{E}(ee)}$, $\rdm{\phi_{A_1}(ee)}{\hat{V}_{ee}}{\phi_{A_1}(ee)}$. The difference in the singlet diagonal matrix elements is
\begin{eqnarray}
\epsilon = \rdm{\phi_{A_1}(ee)}{\hat{V}_{ee}}{\phi_{A_1}(ee)}-\rdm{\phi_{E}(ee)}{\hat{V}_{ee}}{\phi_{E}(ee)} \nonumber \\
= 2\me{e_xe_x}{\hat{V}_{ee}}{e_ye_y}
=2\int \rho_{xy}(\vec{r}_1)\hat{V}_{ee}(\vec{r}_1,\vec{r}_2)\rho_{xy}(\vec{r}_2)d\vec{r}_1d\vec{r}_2\label{eq:singletdif}
\end{eqnarray}
where $\rho_{xy}(\vec{r}) = e_x(\vec{r})e_y(\vec{r})$. The above integral is a standard exchange integral of a charge distribution $\rho(\vec{r})$, which has been proven to be positive definite. One such proof \cite{proof} from electrostatics uses Green's identity to show that
\begin{eqnarray}
\int \rho(\vec{r}_1)\frac{1}{|\vec{r}_1-\vec{r}_2|}\rho(\vec{r}_2)d\vec{r}_1d\vec{r}_2 = \frac{1}{4\pi}\int|\vec{E}(\vec{r})|^2d\vec{r}\geq 0
\end{eqnarray}
where $\vec{E}(\vec{r})$ is the electric field generated by the finite charge distribution $\rho(\vec{r})$.
Thus, considering just the diagonal Coulomb matrix elements, the $^1A_1$ singlet must be higher in energy than the $^1E$ singlet. It has been shown that the difference in the diagonal Coulomb matrix elements of the $^1E$ singlet and $^3A_2$ triplet ($\rdm{\phi_{E}(ee)}{\hat{V}_{ee}}{\phi_{E}(ee)}-\rdm{\phi_{A_2}(ee)}{\hat{V}_{ee}}{\phi_{A_2}(ee)}$) is equal to the difference between the two singlets $\epsilon$ \cite{maze}, thereby confirming the ordering indicated by Hund's rules and implying that the states of the ground configuration are equally spaced prior to coupling with states of the higher MO configurations. In the first excited MO configuration, Hund's rules also indicate that the $^3E$ triplet has a smaller repulsion energy than the $^1E'$ singlet and a similar argument as used for the singlet ordering in the ground MO configuration has been shown to confirm this ordering \cite{paul}.

Since the only configuration states of the ground and first excited MO configurations that have both the same orbital symmetry and spin state are the $^1E$ and $^1E'$ singlets, these are the only states that are coupled by the Coulomb repulsion potential, and all of the other states are solutions of the orbital components of $\hat{H}_{NV}$, $\hat{H}_o = \sum_i\hat{h}_i+\sum_{i>j}\hat{V}_{ee}(\mathbf{x}_i,\mathbf{x}_j)$. The coupled $E$ singlet states $\Phi_{j,k;0,0}^{c'}$ can be expressed as
\begin{eqnarray}
\Phi_{E,k;0,0}^{c'} = N_\kappa[\Phi_{E,k;0,0}^c-\kappa\Phi_{E',k;0,0}^c] \nonumber \\
\Phi_{E',k;0,0}^{c'} = N_\kappa[\Phi_{E',k;0,0}^c+\kappa\Phi_{E,k;0,0}^c]
\end{eqnarray}
where $k=x,y$, the coupling coefficient $\kappa$ is a function of the Coulomb repulsion matrix element
\begin{eqnarray}
\me{\Phi_{E,k;0,0}^c}{\hat{V}_{ee}}{\Phi_{E',k;0,0}^c} = \rdm{\phi_E(a_1e)}{V_{ee}}{\phi_E(ee)}
\end{eqnarray}
and $N_\kappa = \left(1+|\kappa|^2\right)^{-1/2}$ is the normalization constant. The interaction of the two $E$ singlet states will also force the singlets apart in energy, shifting the lower $^1E$ singlet lower in energy towards the ground triplet state, and shifting the higher $^1E'$ singlet further higher in energy than the excited triplet state.

Defining $E_{j;S}$ to be the orbital energies of the configuration states, the derived orbital structure prior to the introduction of spin-orbit and spin-spin interactions is depicted in \fref{fig:orbitalstructure}. The known optical ZPL (1.945 eV) and infrared ZPL (1.190 eV) transition energies are also included in the figure. Notably, the energy separations of the triplet and singlet states have not yet been observed. As indicated by \eref{eq:oneelectronenergies}, all of the configuration states of the ground MO configuration have the same nuclear equilibrium coordinates (ignoring the Coulomb coupling of $^1E$ and $^1E'$). Likewise, all of the configuration states of the first excited MO configuration also have the same nuclear equilibrium coordinates, but these differ from those of the ground MO configuration.  Defining the energy of the ground $^3A_2$ state to be zero ($E_{A_2;1}=0$), the energy of the $^1A_1$ singlet can be expressed as $E_{A_1;0}\approx E_{E;0}+1.190$ eV, where the infrared ZPL has been directly used since, correct to first-order in $\kappa$, both $^1A_1$ and $^1E$ have the same nuclear equilibrium coordinates. The energy of the excited $^3E$ triplet in the nuclear equilibrium coordinates of the ground MO configuration is $E_{E;1} = E_{S}+1.945 \approx 2.180$ eV, where $E_{S} \approx 0.235$ eV is the Stokes shift of the optical transition \cite{gali2}. Thus, the configuration energies $E_{E;0}$ and $E_{E';0}$ of the $E$ singlet states, alongside their Coulomb coupling coefficient $\kappa$, are the first unknown parameters of the molecular model.

A recent \textit{ab initio} study \cite{gali3} concluded that the upper $^1E'$ singlet energetically crossed the $^3E$ triplet as the nearest neighbour carbon and nitrogen nuclei were symmetrically displaced. The proceeding argument clearly shows that this can only be the case if there existed a strong Coulomb repulsion interaction between the $^1E'$ singlet and a higher energy $E$ symmetric singlet. Such an interaction would force the $^1E'$ singlet lower in energy and if it were large enough, it could potentially overcome the repulsion from the lower $^1E$ singlet and the difference in the Coulomb energies between the $^1E'$ singlet and the $^3E$ triplet. The higher energy $E$ singlet that produces this effect is most likely the $^1E''$ singlet that arises from the second excited MO configuration $a_1'a_1^2e^3$. It is also possible that the higher energy $E$ singlet arises from conduction band states, however, it would not be expected that a singlet formed from such delocalized states would interact so strongly with the highly localized states of the centre.  The other relevant second excited MO configuration $a_1'^2e^4$ forms an $^1A_1'$ singlet. Consequently, the singlets of these two second excited MO configurations could potentially force the $E$ and $A_1$ singlets of the ground and first excited MO configurations down in energy by different amounts, resulting in the $^1A_1$ crossing one of the $E$ singlets. The most recent \textit{ab initio} studies \cite{delaney, gali3} indicate that the ordering of the ground MO configuration singlets depicted in \fref{fig:orbitalstructure} is correct and without further evidence that the second excited MO configurations states significantly shift the positions of the lower singlets, the simple orbital structure of \fref{fig:orbitalstructure} will be assumed. Due to the parametric formulation of the molecular model, the failure of this assumption will not effect the validity of the expressions derived in this analysis.

Another recent \textit{ab initio} study \cite{delaney} indicates that the $^1A_1$ singlet is close to and possibly higher in energy than the $^3E$ triplet. As the quantitative values of the Coulomb repulsion matrix elements are not known, the proceeding analysis can not offer a definitive conclusion on the ordering of the $^1A_1$ singlet in relation to the $^3E$ triplet. The current understanding of the mechanisms of spin polarization and optical dynamics \cite{manson, singletlevel} appears to strongly indicate that a singlet state is close, but lower in energy than the $^3E$ triplet. Therefore, further \textit{ab initio} work is required to definitively determine the ordering of the $^1A_1$, $^3E$ and $^1E'$ states.

Thus far, it has been shown how the orbital structure of the NV$^-$ centre can be derived from the definition of the centre's MOs. The configuration states that have been obtained are solutions of the orbital components $\hat{H}_o$ of the centre's Hamiltonian and the orbital structure was determined up to the two unknown energies of the $E$ symmetric singlets. The analysis of the Coulomb repulsion matrix elements has offered an insight into the contention surrounding the ordering of the intermediate singlet states. In the next section, the spin-orbit $\hat{V}_{so}$ and electronic spin-spin $\hat{V}_{ss}$ potentials will be treated as first-order perturbations to $\hat{H}_o$ and the fine structure and mixing of electronic spin of the centre will be determined.

\section{Fine structure and the mixing of electronic spin}

The construction of the configuration states in the previous section to have well defined orbital symmetry aided in simplifying the treatment of the orbital components $\hat{H}_o$ of the centre's Hamiltonian. Since spin-orbit and spin-spin interactions act on both the electronic orbital and spin coordinates, their treatment can be likewise greatly simplified by constructing linear combinations of the configuration states that have well defined spin-orbit symmetry. This can be conducted by first defining the combinations of the S=1,0 spin states ($\ket{S,m_s}$) that transform as particular rows of irreducible representations of the $C_{3v}$ group \cite{lenef}
\begin{eqnarray}
S_{A_1} = \ket{0,0}, \ \ S_{A_2} = \ket{1,0}, \nonumber \\
S_{E,x} = \frac{-i}{\sqrt{2}}(\ket{1,1}+\ket{1,-1}), \ \ S_{E,y} = \frac{-1}{\sqrt{2}}(\ket{1,1}-\ket{1,-1})
\label{eq:spinstates}
\end{eqnarray}
In an analogous method to that used in constructing the combinations of products of two MOs that had well defined symmetry, the above symmeterized spin states can be used in conjunction with the configuration states to construct the symmeterized spin-orbit states $\Phi_{n,j,k}^{so}$ contained in \tref{tab:states}, where $n$ denotes the energy level, and $j$ and $k$ denote the irreducible representation and row that the spin-orbit state transforms as in spin-orbit space.

The spin-orbit and spin-spin interaction potentials are \cite{tinkham}
\begin{eqnarray}
\hat{V}_{so} = \frac{1}{2m^2c^2}\sum_i\vec{\nabla}\hat{V}_{Ne}(\vec{r}_i)\times\vec{p}_i\cdot\vec{s}_i \equiv \sum_i\vec{\lambda}_i\cdot\vec{\sigma}_i \nonumber \\
\hat{V}_{ss} = \frac{\mu_0g_e^2\mu_B^2}{4\pi\hbar^2}\sum_{i>j}\frac{\vec{s}_i\cdot\vec{s}_j}{|\vec{r}_{ij}|^3}
-\frac{3(\vec{s}_i\cdot\vec{r}_{ij})(\vec{r}_{ij}\cdot\vec{s}_j)}{|\vec{r}_{ij}|^5}
\equiv \sum_{i>j}\vec{\sigma}_i\cdot\bar{D}_{i,j}\cdot\vec{\sigma}_j
\end{eqnarray}
where $\vec{p}_i$ and $\vec{s}_i = (\hbar/2)\vec{\sigma}_i$ are the momentum and spin operators of the $i^{th}$ electron, $\vec{r}_{ij} = \vec{r}_j-\vec{r}_i = x_{ij}\vec{x}+y_{ij}\vec{y}+z_{ij}\vec{z}$ ($\vec{x},\vec{y},\vec{z}$ unit coordinate vectors), $g_e$ is the free electron g-factor, and $\vec{\lambda}$ and $\bar{D}$ are rank one and two orbital tensor operators respectively. The components of the orbital tensor operators are
\begin{eqnarray}
\vec{\lambda} = -\hat{\lambda}_{E,y}\vec{x}+\hat{\lambda}_{E,x}\vec{y}+\hat{\lambda}_{A_2}\vec{z} \nonumber \\
\bar{D} = \left(\begin{array}{ccc}
-\frac{1}{2}\hat{D}_{A_1}-\hat{D}_{E,x,1} & \hat{D}_{E,y,1} & -\hat{D}_{E,x,2} \\
\hat{D}_{E,y,1} & -\frac{1}{2}\hat{D}_{A_1}+\hat{D}_{E,x,1} & -\hat{D}_{E,y,2} \\
-\hat{D}_{E,x,2} & -\hat{D}_{E,y,2} & \hat{D}_{A_1} \\
\end{array}\right)
\end{eqnarray}
where
\begin{eqnarray}
\hat{D}_{A_1} = \frac{\mu_0g_e^2\mu_B^2}{16\pi|\vec{r}_{ij}|^3}(1-\frac{3z_{ij}^2}{|\vec{r}_{ij}|^5}), \nonumber \\
\hat{D}_{E,x,1} = \frac{\mu_0g_e^2\mu_B^2}{32\pi|\vec{r}_{ij}|^3}\frac{3(x_{ij}^2-y_{ij}^2)}{|\vec{r}_{ij}|^5}, \ \
\hat{D}_{E,y,1} = -\frac{\mu_0g_e^2\mu_B^2}{32\pi|\vec{r}_{ij}|^3}\frac{6x_{ij}y_{ij}}{|\vec{r}_{ij}|^5}, \nonumber \\
\hat{D}_{E,x,1} = \frac{\mu_0g_e^2\mu_B^2}{16\pi|\vec{r}_{ij}|^3}\frac{3z_{ij}x_{ij}}{|\vec{r}_{ij}|^5}, \ \
\hat{D}_{E,y,1} = \frac{\mu_0g_e^2\mu_B^2}{16\pi|\vec{r}_{ij}|^3}\frac{3z_{ij}y_{ij}}{|\vec{r}_{ij}|^5} \nonumber
\end{eqnarray}
Using the spin-orbit states of \tref{tab:states} and the above definitions of the tensor operators, the application of the Wigner-Eckart theorem allows the computation of the matrix representations of the spin-orbit and spin-spin potentials in terms of one- and two-electron reduced density matrix elements. The matrix representations are contained in tables \ref{tab:SOmatrix} and \ref{tab:SSmatrix} and the reduced density matrix element expressions are contained in \tref{tab:parameters}. Note that as only the spin-orbit states associated with the triplet configuration states have non-zero spin-spin matrix elements, the matrix representation of the spin-spin potential is presented in the reduced basis of triplet spin-orbit states. In order to maintain clarity, only the upper triangle of the matrix representations have been presented. The lower half can be inferred using the hermitian property of the potentials and the fact that each of the reduced density matrix element expressions are real.

\Table{\label{tab:SOmatrix}The matrix representation of the spin-orbit interaction potential in the basis $\{\Phi_{1,A_1}^{so},\Phi_{2,E,x}^{so},\Phi_{2,E,y}^{so},\Phi_{3,E,x}^{so},\Phi_{3,E,y}^{so},\Phi_{4,A_1}^{so},\Phi_{5,E,x}^{so},\Phi_{5,E,y}^{so},\Phi_{6,E,x}^{so},\Phi_{6,E,y}^{so},
\Phi_{7,A_2}^{so},\Phi_{8,A_1}^{so},\Phi_{9,E,x}^{so},\Phi_{9,E,y}^{so}\}$. The spin-orbit parameters $\lambda_\parallel$ and $\lambda_\perp$ are defined in \tref{tab:parameters}. The lower half of the matrix can be obtained using the hermitian property of the spin-orbit potential.}
\br
0   &   0   &   0   &   0   &   0   &   $-2i\lambda_\parallel $   &   0    &   0    &   0     &   0     &   0   &   $\sqrt{2}\lambda_\perp$ & 0 & 0  \\
    &   0   &   0   &   0   &   0   &   0       &   0    &   0    &   $-\lambda_\perp$   &   0     &   0   &   0   & $-i\lambda_\perp$ & 0 \\
    &       &   0   &   0   &   0   &   0       &   0    &   0    &   0     &   $-\lambda_\perp$   &   0   &   0   & 0 & $-i\lambda_\perp$ \\
    &       &       &   0   &   0   &   0       &   $-i\sqrt{2}\lambda_\perp$    &   0    &   0     &   0     &   0   &   0  & 0 & 0  \\
    &       &       &       &   0   &   0       &   0    &   $-i\sqrt{2}\lambda_\perp$    &   0     &   0     &   0   &   0  & 0 & 0  \\
    &       &       &       &       &   0       &   0    &   0    &   0     &   0     &   0   &   $i\sqrt{2}\lambda_\perp$  & 0 & 0  \\
    &       &       &       &       &           &  $ -\lambda_\parallel$   &   0    &   0     &   0     &   0   &   0   & 0 & 0 \\
    &       &       &       &       &           &        &   $-\lambda_\parallel$   &   0     &   0     &   0   &   0   & 0 & 0 \\
    &       &       &       &       &           &        &        &   0     &   0     &   0   &   0   & $-i\lambda_\parallel$ & 0 \\
    &       &       &       &       &           &        &        &         &   0     &   0   &   0   & 0 & $-i\lambda_\parallel$ \\
    &       &       &       &       &           &        &        &         &         &   $\lambda_\parallel$   &   0    & 0 & 0\\
    &       &       &       &       &           &        &        &         &         &       &   $\lambda_\parallel$    & 0 & 0\\
        &       &       &       &       &           &        &        &         &         &       &       & 0 & 0\\
            &       &       &       &       &           &        &        &         &         &       &     &  & 0\\
\br
\endTable

\fulltable{\label{tab:SSmatrix}The matrix representation of spin-spin interaction potential in the triplet basis $\{\Phi_{1,A_1}^{so},\Phi_{2,E,x}^{so},\Phi_{2,E,y}^{so},\Phi_{5,E,x}^{so},\Phi_{5,E,y}^{so},\Phi_{6,E,x}^{so},\Phi_{6,E,y}^{so},
\Phi_{7,A_2}^{so},\Phi_{8,A_1}^{so}\}$.  The spin-spin parameters are defined in \tref{tab:parameters}. The lower half of the matrix can be obtained using the hermitian property of the spin-spin potential.}
\br
$-2D_{1,A_1}$ & 0 & 0  & 0 & 0 & 0 & 0 & 0 & $\sqrt{2}D_{1,E,2}$ \\
           & $D_{1,A_1}$ & 0 &  $\sqrt{2}D_{1,E,1}$ & 0 & $D_{1,E,2}$ & 0 & 0 & 0 \\
           &            & $D_{1,A_1}$ &  0 & $\sqrt{2}D_{1,E,1}$ & 0 & $D_{1,E,2}$ & 0 & 0  \\
           &            &             & $D_{2,A_1}$ & 0 & $\sqrt{2}D_{2,E,2}$ & 0 & 0 & 0  \\
           &            &             &   & $D_{2,A_1}$ & 0 & $\sqrt{2}D_{2,E,2}$ & 0 & 0  \\
           &            &             &   &   & $-2D_{2,A_1}$ & 0 & 0 & 0  \\
           &            &             &   &   &   & $-2D_{2,A_1}$ & 0 & 0  \\
           &            &             &   &   &   &   & $D_{2,A_1}-2D_{2,E,1}$ & 0  \\
           &            &             &   &   &   &   &  & $D_{2,A_1}+2D_{2,E,1}$  \\
\br
\endfulltable

In calculating the non-axial orbital components of the spin-orbit potential, it was concluded that the reduced density matrix element $\rdm{e}{\lambda_E}{e}$ must be zero and so it was not included in the matrix representation of \tref{tab:SOmatrix}. This follows from the fact that as $\lambda_{E,x}$ is a purely imaginary hermitian orbital operator and its diagonal matrix elements in the MO basis are $\me{e_x}{\lambda_{E,x}}{e_x} = \frac{1}{\sqrt{2}}\rdm{e}{\lambda_E}{e} = -\me{e_y}{\lambda_{E,x}}{e_y}$, these diagonal matrix elements must vanish, implying that $\rdm{e}{\lambda_E}{e}=0$. In a previous application of the molecular model to the fine structure of the excited triplet state \cite{tamarat} this conclusion was not made and the coupling of $\Phi_{5,E,k}^{so}$ and $\Phi_{6,E,k}^{so}$ was incorrectly assigned as arising from the non-axial spin-orbit interaction matrix element $-i\sqrt{2}\rdm{e}{\lambda_E}{e}$ instead of the correct spin-spin interaction matrix element $\sqrt{2}D_{2,E,2}$ as contained in \tref{tab:SSmatrix}.

\Table{\label{tab:parameters}The spin-orbit and spin-spin parameters of the molecular model expressed in terms of one- and two-electron reduced density matrix elements containing the molecular orbitals. Known values \cite{excitedstatestrain} and estimated order of magnitude of unknown values are contained in the right most column.}
\br
 Parameter & Expression & Value \\
\mr
$\lambda_\parallel$ & $-i\rdm{e}{\lambda_{A_2}}{e}$ & 5.3 GHz \\
$\lambda_\perp$ & $\frac{-i}{\sqrt{2}}\rdm{a_1}{\lambda_{E}}{e}$ & $\sim$ GHz \\
$D_{1,A_1}$ & $2\rdm{\phi_{A_2}(ee)}{D_{A_1}}{\phi_{A_2}(ee)}$ & 2.87/3 GHz \\
$D_{1,E,1}$ & $4\rdm{\phi_{E}(ae)}{D_{E,1}}{\phi_{A_2}(ee)}$ & $\sim$ MHz \\
$D_{1,E,2}$ & $-4\rdm{\phi_{E}(ae)}{D_{E,2}}{\phi_{A_2}(ee)}$ & $\sim$ MHz \\
$D_{2,A_1}$ & $\rdm{\phi_{E}(ae)}{D_{A_1}}{\phi_{E}(ae)}-\rdm{\phi_{E}(ae)}{D_{A_1}}{\phi_{E}(ea)}$ & 1.42/3 GHz \\
$D_{2,E,1}$ & $-2(\rdm{\phi_{E}(ae)}{D_{E,1}}{\phi_{E}(ae)}-\rdm{\phi_{E}(ae)}{D_{E,1}}{\phi_{E}(ea)})$ & 1.55/2 GHz \\
$D_{2,E,2}$ & $2(\rdm{\phi_{E}(ae)}{D_{E,2}}{\phi_{E}(ae)}-\rdm{\phi_{E}(ae)}{D_{E,2}}{\phi_{E}(ea)})$ & $200/\sqrt{2}$ MHz \\
\br
\endTable

\begin{figure}[hbtp]
\begin{center}
\mbox{
\subfigure[]{\includegraphics[width=0.275\columnwidth] {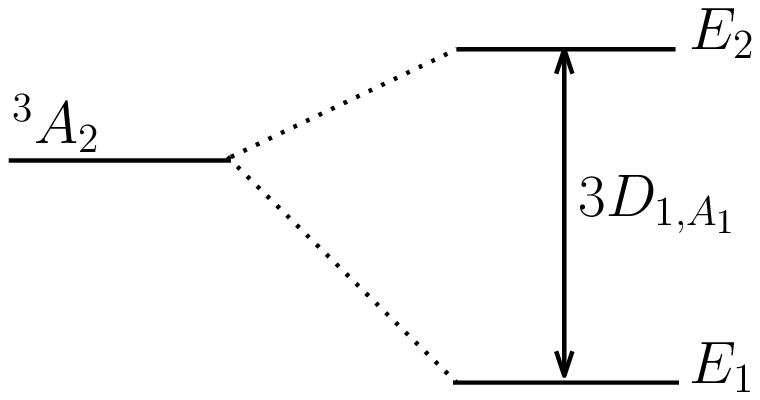}}
\subfigure[]{\includegraphics[width=0.33\columnwidth] {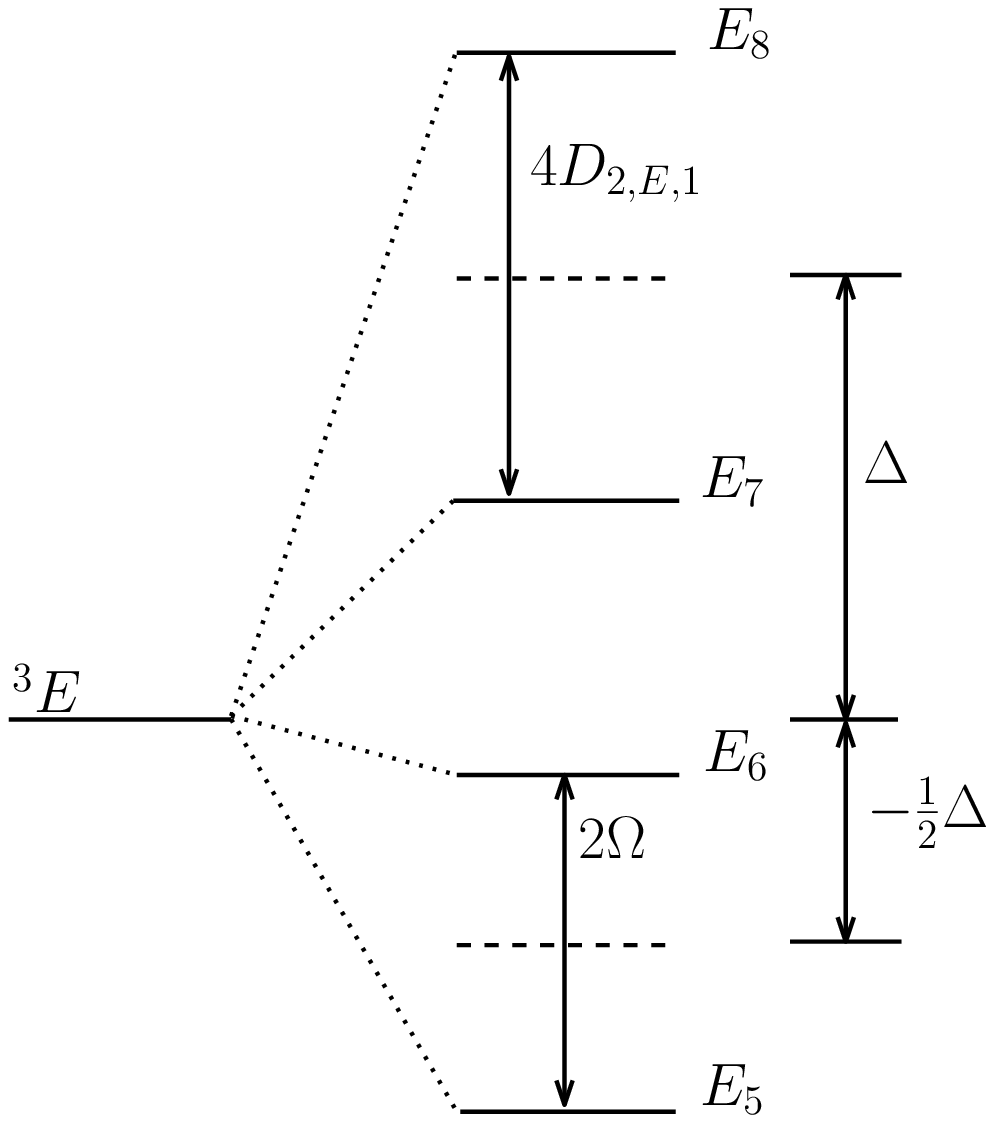}}
\subfigure[]{\includegraphics[width=0.3\columnwidth] {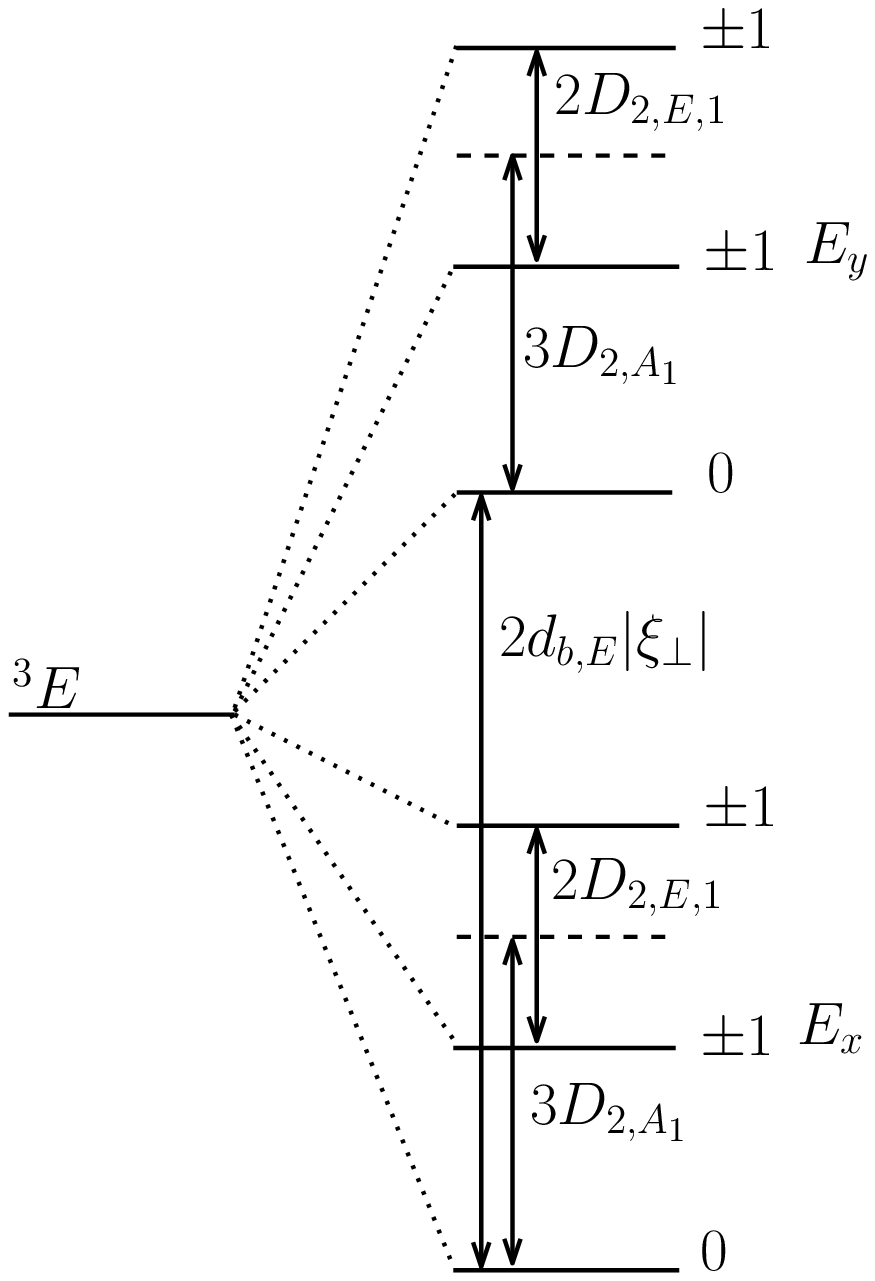}}
}
\caption{Energy level diagrams of the fine structure of the $^3A_2$ and $^3E$ triplets of the NV$^-$ centre: (a) $^3A_2$ fine structure independent of strain; (b) $^3E$ fine structure in the absence of strain; and, (c) $^3E$ fine structure in the limit of large non-axial strain $|\xi_\perp|$. Expressions of the energies $E_n$ are contained in \tref{tab:energies} and $\Delta = \lambda_\parallel+D_{2,A_1}$ and $\Omega = \frac{1}{2}[(\lambda_\parallel-3D_{2,A_1})^2+8|D_{2,E,2}|^2]^{1/2}$. The strain interaction parameter $d_{b,E}|\xi_\perp|$ is defined in \eref{eq:sostrain}.}
\label{fig:finestructure}
\end{center}
\end{figure}

As the spin-orbit and spin-spin matrix elements are observed to be of the order of MHz-GHz \cite{tamarat, excitedstatestrain} and the energy separations of the singlet and triplet states are expected to be of the order of meV - eV ($\sim10^2-10^5$ GHz), it is appropriate to treat the spin-orbit and spin-spin potentials together as first order perturbations to $\hat{H}_o$. The configuration energies $E_{j;S}$ and spin-orbit states $\{\Phi_{n,j,k}^{so}\}$ are then the zero order energies and electronic states of the perturbation expansion. The application of first order perturbation theory yields the fine structure energies $E_n$ contained in \tref{tab:energies} and depicted in \fref{fig:finestructure} as well as the coupling coefficients contained in \tref{tab:couplingcoef}. The first order corrected electronic states $\Phi_{n,j,k}^{so'}$ are defined in terms of the coupling coefficients by
\begin{eqnarray}
\Phi_{n,j,k}^{so'} = N_n[s_{n,n}^{(0)}\Phi_{n,j,k}^{so}+\sum_m(s_{n,m}^{(1)}+s_{n,m}^{(2)})\Phi_{m,j,k}^{so}]
\end{eqnarray}
where $N_n$ is the normalization constant. The coefficient $\eta$ contained in \tref{tab:couplingcoef} is the coefficient that arises from the coupling of the degenerate $\Phi_{5,E,k}^{so}$ and $\Phi_{6,E,k}^{so}$ states of the $^3E$ triplet due to spin-spin interaction and is expressed in terms of the spin-orbit and spin-spin parameters as
\begin{eqnarray}
\eta = \frac{2\sqrt{2}D_{2,E,2}}{\lambda_\parallel-3D_{2,A_1}+\left[(\lambda_\parallel-3D_{2,A_1})^2+8|D_{2,E,2}|^2\right]^{\frac{1}{2}}}
\end{eqnarray}
and $N_\eta = \left(1+|\eta|^2\right)^{-1/2}$ is the associated normalization constant. Note that the coefficients $s_{n,m}^{(2)}$ do not denote coefficients that are truly second order in the spin-orbit and spin-spin parameters, but rather contain the first order products of the Coulomb $\kappa$ and degenerate spin-spin $\eta$ coefficients with the other spin-orbit and spin-spin coefficients. As $\kappa$ and $\eta$ are expected to be orders of magnitude larger than the other spin-orbit and spin-spin coefficients, the $s_{n,m}^{(2)}$ coefficients may potentially be only slightly smaller than the $s_{n,m}^{(1)}$ coefficients.

\Table{\label{tab:energies}The electronic energies correct to first order in spin-orbit and spin-spin interactions. The energies calculated using the known parameters of \tref{tab:parameters} are provided in the right column.}
\br
 $E_n$ & & $E_n^{(0)}$ & $E_n^{(1)}$ & Calc. \\
 \mr
$E_1$ & = & $E_{A_2;1}$ & $-2D_{1,A_1}$ & -1.91 GHz \\
$E_2$ & = & $E_{A_2;1}$ & $+D_{1,A_1}$ & 0.957 GHz\\
$E_3$ & = & $E_{E;0}$ & - & - \\
$E_4$ & = & $E_{A_1;0}$ & - & - \\
$E_5$ & = & $E_{E;1}$ & $-\frac{1}{2}(\lambda_\parallel+D_{2,A_1})-\frac{1}{2}\left[(\lambda_\parallel-3D_{2,A_1})^2+8|D_{2,E,2}|^2\right]^{\frac{1}{2}}$ & $E_{E;1}$ - 4.84 GHz\\
$E_6$ & = & $E_{E;1}$ & $-\frac{1}{2}(\lambda_\parallel+D_{2,A_1})+\frac{1}{2}\left[(\lambda_\parallel-3D_{2,A_1})^2+8|D_{2,E,2}|^2\right]^{\frac{1}{2}}$ & $E_{E;1}$ - 0.936 GHz\\
$E_7$ & = & $E_{E;1}$ & $+\lambda_\parallel+D_{2,A_1}-2D_{2,E,1}$ & $E_{E;1}$ +4.22 GHz\\
$E_8$ & = & $E_{E;1}$ & $+\lambda_\parallel+D_{2,A_1}+2D_{2,E,1}$ & $E_{E;1}$ +7.32 GHz\\
$E_9$ & = & $E_{E';0}$ & - & - \\
\br
\endTable

The spin-orbit and spin-spin parameters that are contained in the first order energies of \tref{tab:energies} have been determined in previous studies \cite{tamarat, excitedstatestrain} by directly observing the fine structure of the ground and excited triplet states and modeling the variation of the excited triplet fine structure with the application of strain and electric fields at low temperatures. The values of these known parameters are contained alongside the unknown parameters in \tref{tab:parameters}. The unknown parameters are all involved only in the first order couplings of the spin-orbit states of the singlets and triplets and do not contribute to the first order energies. Consequently, the effect of these unknown parameters can only be indirectly detected through the observation of interactions of the centre with electric, magnetic and strain perturbations that can not be explained by the zero order spin-orbit states and the coupling coefficients arising from the known parameters. Examples of such interactions are the presence of the Stark effect in the ground state triplet \cite{efield,vanoort} and the anisotropy of the ground state effective g-factor \cite{gfactor}. Frustratingly, it is precisely these unknown parameters which are likely to govern the process of spin-polarization as they allow forbidden transitions between the triplet and singlet states, however, this will not be discussed further in this article. Using the known values of similar parameters, the estimated order of magnitude of the unknown parameters are also contained in \tref{tab:parameters}.

\section{Electric, magnetic and strain interactions}

The Stark shift $\hat{V}_S$, Zeeman effect $\hat{V}_Z$ and strain $\hat{V}_\xi$ potentials that describe the centre's interaction with electric $\vec{E}$, magnetic $\vec{B}$ and effective strain $\vec{\xi}$ fields are
\begin{eqnarray}
\hat{V}_{S} = \sum_i\vec{d}_i\cdot\vec{E}, \ \ \hat{V}_Z = \frac{\mu_B}{\hbar}\sum_i(\vec{l}_i+g_e\vec{s}_i)\cdot\vec{B}, \ \
\hat{V}_\xi = \sum_i\vec{d}_i\cdot\vec{\xi}
\end{eqnarray}
where $\vec{d}_i = e\vec{r}_i$ is the electric dipole operator and $\vec{l}_i = \vec{r}_i\times\vec{p}_i$ is the orbital magnetic moment operator. Note that the above effective expression for the strain potential (in which strain is treated as an effective local electric field) is derived using the group operator replacement theorem \cite{cornwell} and can be unambiguously interpreted for uniaxial strain, but some care is required in its interpretation for non-uniaxial strain.

Each of these potentials contain operators that act on just the electronic orbital coordinates and can be written in terms of sums of orbital tensor operators with definite symmetry properties
\begin{eqnarray}
\vec{d}_i = \hat{d}_{E,x}\vec{x}+\hat{d}_{E,y}\vec{y}+\hat{d}_{A_1}\vec{z}, \ \
\vec{l}_i = -\hat{l}_{E,y}\vec{x}+\hat{l}_{E,x}\vec{y}+\hat{l}_{A_2}\vec{z}
\end{eqnarray}
As the spin-orbit states constructed in the last section are simple linear combinations of configuration states that have both well defined orbital symmetry and spin, it is straight forward to apply the Wigner-Eckart theorem to calculate the matrix representations of each of the tensor orbital operators in terms of a set of reduced density matrix elements. Since the expressions of the reduced density matrix elements are identical for all orbital tensor operators of the same symmetry, the matrix representations of general orbital tensor operators $\hat{O} = \sum_i\hat{O}_{i}$ of each symmetry are contained in tables \ref{tab:orbitaloperators1} and \ref{tab:orbitaloperators2}. The matrix representation of the electron spin operator $\vec{S}=\sum_i\vec{s}_i$ is contained in table \ref{tab:spinoperators}. These matrix representations can be immediately applied to calculate the effects of the different electric, magnetic and strain interactions in the basis of the zero order spin-orbit states.

\begin{figure}[hbtp]
\begin{center}
\includegraphics[width=0.7\columnwidth] {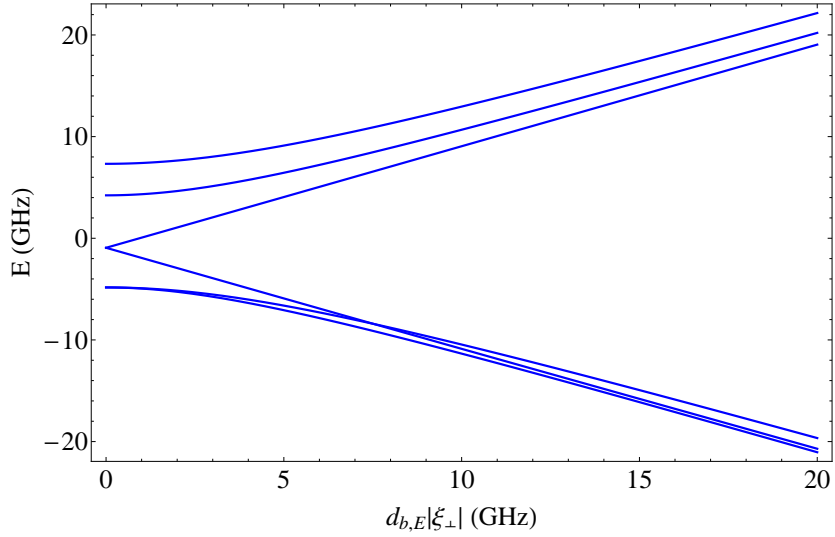}
\caption{Plot of the variation of the $^3E$ triplet fine structure with the magnitude of the applied non-axial strain $|\xi_\perp|$. The plot was produced using the known spin-orbit and spin-spin parameters of \tref{tab:parameters} and the matrix representation of the strain interaction \eref{eq:sostrain}.}
\label{fig:strainplot}
\end{center}
\end{figure}

For example, the matrix representation of the strain potential in the basis of the zero order spin-orbit states of the $^3E$ triplet $\{\Phi_{5,E,x}^{so},\Phi_{5,E,y}^{so},\Phi_{6,E,x}^{so},\Phi_{6,E,y}^{so},\Phi_{7,A_2}^{so},\Phi_{8,A_1}^{so}\}$ is
\begin{eqnarray}
V_\xi[^3E] = \nonumber \\
\left(\begin{array}{cccccc}
d_{b,A_1}\xi_z & 0 & 0 & 0 & -d_{b,E}\xi_y & -d_{b,E}\xi_x \\
0 & d_{b,A_1}\xi_z & 0 & 0 & d_{b,E}\xi_x & -d_{b,E}\xi_y \\
0 & 0 & d_{b,A_1}\xi_z+d_{b,E}\xi_x & -d_{b,E}\xi_y & 0 & 0 \\
0 & 0 & -d_{b,E}\xi_y & d_{b,A_1}\xi_z-d_{b,E}\xi_x  & 0 & 0 \\
-d_{b,E}\xi_y & d_{b,E}\xi_x & 0 & 0 & d_{b,A_1}\xi_z & 0 \\
-d_{b,E}\xi_x & -d_{b,E}\xi_y & 0 & 0 & 0 & d_{b,A_1}\xi_z  \\
\end{array}
\right) \nonumber \\
\label{eq:sostrain}
\end{eqnarray}
where $d_{b,A_1} = \rdm{a_1}{d_{A_1}}{a_1}+3\rdm{e}{d_{A_1}}{e}$ and $d_{b,E} = \rdm{e}{d_E}{e}$. Previous studies \cite{tamarat,excitedstatestrain} have shown that the above matrix representation yields a strain variation of the $^3E$ triplet fine structure that agrees excellently with observation. A theoretical plot of the effect of non-axial strain is depicted in \fref{fig:strainplot} and it clearly shows that non-axial strain splits the $^3E$ fine structure into upper and lower branches.  Due to the $D_{2,E,2}$ spin-spin parameter that couples the $m_s = 0$ and $ m_s = \pm1$ configuration states, there exists two level anti-crossings in the lower branch as the $m_s = 0$ changes from being the highest energy state of the lower branch to the lowest energy state \cite{tamarat}. In the vicinity of these anti-crossings the states of the lower branch are all significant mixtures of the different spin sub-levels \cite{tamarat}. As there is no anti-crossing in the upper branch, the mixing of the spin sub-levels in the states of the upper branch does not significantly vary with strain \cite{tamarat}.

In the limit of large strain the upper and lower branches become identical, with the same mixing and splitting of the spin sub-levels. This can be demonstrated by calculating the matrix representations of the spin-orbit, spin-spin and strain potentials in the basis of the $^3E$ configuration states $\{\Phi_{E,x;1,0}^c,\Phi_{E,x;1,1}^c,\Phi_{E,x;1,-1}^c,\Phi_{E,y;1,0}^c,\Phi_{E,y;1,1}^c,\Phi_{E,y;1,-1}^c\}$
\begin{eqnarray}
V_{so}[^3E]+V_{ss}[^3E] = \nonumber\\
\left(\begin{array}{ccc:ccc}
-2D_{2,A_1} & \frac{D_{2,E,2}}{\sqrt{2}} & -\frac{D_{2,E,2}}{\sqrt{2}} & 0 & \frac{-D_{2,E,2}}{\sqrt{2}} & \frac{-D_{2,E,2}}{\sqrt{2}} \\
\frac{D_{2,E,2}}{\sqrt{2}} & D_{2,A_1} & D_{2,E,1} & \frac{D_{2,E,2}}{\sqrt{2}} & \lambda_\parallel & -D_{2,E,1} \\
\frac{-D_{2,E,2}}{\sqrt{2}} & D_{2,E,1} & D_{2,A_1} & \frac{D_{2,E,2}}{\sqrt{2}} & D_{2,E,1} & -\lambda_\parallel \\
\hdashline
0 & \frac{D_{2,E,2}}{\sqrt{2}} & \frac{D_{2,E,2}}{\sqrt{2}} & -2D_{2,A_1}  & \frac{-D_{2,E,2}}{\sqrt{2}}& \frac{D_{2,E,2}}{\sqrt{2}} \\
\frac{-D_{2,E,2}}{\sqrt{2}} & \lambda_\parallel & D_{2,E,1} & \frac{-D_{2,E,2}}{\sqrt{2}} & D_{2,A_1} & -D_{2,E,1} \\
\frac{-D_{2,E,2}}{\sqrt{2}} & -D_{2,E,1} & -\lambda_\parallel & \frac{D_{2,E,2}}{\sqrt{2}} & -D_{2,E,1} & D_{2,A_1}  \\
\end{array}
\right) \nonumber
\end{eqnarray}
\begin{eqnarray}
V_\xi[^3E] =
\left(\begin{array}{ccc:ccc}
-d_{b,E}\xi_x & 0 & 0 & d_{b,E}\xi_y & 0 & 0 \\
0 & -d_{b,E}\xi_x & 0 & 0 & d_{b,E}\xi_y & 0 \\
0 & 0 & -d_{b,E}\xi_x & 0 & 0 & d_{b,E}\xi_y \\
\hdashline
d_{b,E}\xi_y & 0 & 0 & d_{b,E}\xi_x  & 0 & 0 \\
0 & d_{b,E}\xi_y & 0 & 0 & d_{b,E}\xi_x & 0 \\
0 & 0 & d_{b,E}\xi_y & 0 & 0 & d_{b,E}\xi_x  \\
\end{array}
\right)\label{eq:configstrain}
\end{eqnarray}
Clearly, if a $\xi_x$ strain was applied such that $d_{b,E}\xi_x$ was much larger than the spin-orbit and spin-spin parameters, then the influence of the matrix elements of the upper right and lower left blocks would become insignificant and the configuration states would separate into identical $E_x$ and $E_y$ orbital branches. The off diagonal spin-spin matrix elements in the diagonal blocks are responsible for the mixing and splitting of the spin sub-levels within each branch. The fine structure of the $^3E$ triplet in the high strain limit is also depicted in \fref{fig:finestructure}.

Strain also effects the $^1E'$ singlet and, only after the Coulomb coupling of the $E$ singlets, the $^1E$ singlet as well at first order. The matrix representations of the strain potential correct to first order in $\kappa$ in each of the corresponding basis sets $\{\Phi_{3,E,x}^{c'},\Phi_{3,E,y}^{c'}\}$ and $\{\Phi_{9,E,x}^{c'}, \Phi_{9,E,y}^{c'}\}$ are
\begin{eqnarray}
V_\xi[^1E] = N_\kappa^2\left(\begin{array}{cc}
d_{a,A_1}\xi_z+2\kappa d_{a,E}\xi_x & -2\kappa d_{a,E}\xi_y \\
-2\kappa d_{a,E}\xi_y & d_{a,A_1}\xi_z-2\kappa d_{a,E}\xi_x \\
\end{array}\right) \nonumber \\
V_\xi[^1E'] = N_\kappa^2\left(\begin{array}{cc}
d_{b,A_1}\xi_z-(d_{b,E}+2\kappa d_{a,E})\xi_x & (d_{b,E}+2\kappa d_{a,E})\xi_y \\
(d_{b,E}+2\kappa d_{a,E})\xi_y & d_{b,A_1}\xi_z+(d_{b,E}+2\kappa d_{a,E})\xi_x \\
\end{array}\right) \nonumber \\
 \label{eq:singletstrain}
\end{eqnarray}
where $d_{a,A_1} = 2\rdm{a_1}{d_{A_1}}{a_1}+2\rdm{e}{d_{A_1}}{e}$ and $d_{a,E} = \rdm{a_1}{d_E}{e}$. The strain splitting of the $^1E'$ singlet has not yet been directly observed, however the strain splitting of the $^1E$ singlet has been observed \cite{infrared} in the strain dependence of the infrared transition at 1.190 eV, thereby confirming the coupling of the $E$ symmetric singlets. Investigations of this strain splitting and the information that it can provide about the mixing of the $E$ singlets are on going.

The operator matrix representations constructed in this section may also be used in conjunction with the first order corrected spin-orbit states presented in the previous section to investigate the appearance of interactions at first and higher orders in the spin-orbit and spin-spin parameters. As discussed, these interactions provide important information on the unknown parameters and should be the subject of future investigations.

\section{The room temperature electronic structure}

The dynamic Jahn-Teller effect has been observed in the $^3E$ triplet \cite{jahnteller}. This effect
arises from the vibronic coupling of the spin-orbit states of the triplet \cite{vibcoupling}. The presence of the dynamic Jahn-Teller has the consequence that with increasing temperature $E$ symmetric phonons drive transitions between the states of the triplet. The phonon rates are dependent on temperature and firstly are observed to give an unusual temperature dependence of the optical linewidths and at higher temperatures the phonon transition rates become much greater than the observable lifetimes of the states \cite{stoneham}. At this point, any population which is excited into one of the states is distributed to the other states that it is coupled with before any radiative decay can occur. If the phonon transitions distribute equal population to each of the coupled states then the radiative transition will have the average energy of the coupled states. This process has been previously discussed in the literature as orbital averaging \cite{averaging}.

\begin{figure}[hbtp]
\begin{center}
\includegraphics[width=0.5\columnwidth] {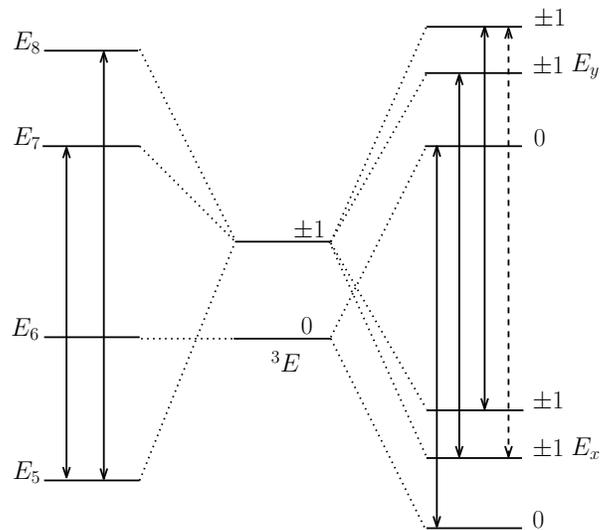}
\caption{Schematics of the fine structure of the $^3E$ triplet at low temperature low strain (left), low temperature high strain (right), and ambient temperature (centre). The solid arrows indicate allowed phonon transitions and the dashed arrow indicates the transition allowed once $D_{2,E,1}$ spin-spin coupling within the two orbital branches in the high strain case is introduced. The plot demonstrates that if there were equal population in each of the phonon coupled states, the fine structure would average to the central structure with a single spin-spin splitting of $3D_{2,A_1}= 1.42$ GHz regardless of strain.}
\label{fig:jtcoupling}
\end{center}
\end{figure}

Since electron-phonon coupling is an orbital operator, the matrix representations \eref{eq:sostrain} and \eref{eq:configstrain} of an orbital operator in the $^3E$ triplet can be used to determine the selection rules of the phonon transitions. The allowed transitions are depicted in \fref{fig:jtcoupling} for both the low and high strain cases. It is clear that for both the low and high strain cases, the phonon transitions will distribute population between the states of a particular spin projection, and since the transition matrix elements are identical for each transition, there will be approximately equal population distributed to each of the states of a given spin sub-level. Due to the coupling of the states with $m_s = \pm1$ spin projections by the $D_{2,E,1}$ spin-spin parameter, the population will also be distributed between these sub-levels. Thus, regardless of strain, the fine structure of the $^3E$ triplet will average to a single splitting of $3D_{2,A_1}= 1.42$ GHz between the $m_s = 0$ and $m_s = \pm1$ spin sub-levels arising from spin-spin interaction (as depicted in \fref{fig:jtcoupling}). Therefore, as observed \cite{averaging}, the fine structure of the $^3E$ triplet at room temperature does not vary between NV$^-$ centres and appears as an effective orbital singlet split by spin-spin, similar to the ground $^3A_2$ triplet.

The matrix representations of orbital operators in the $^1E$ and $^1E'$ singlets \eref{eq:singletstrain} indicate that these states should also be susceptible to the dynamic Jahn-Teller effect, although this has not yet been observed. The Jahn-Teller effect is likely to have significant consequences for the non-radiative dynamics of these states and influence their possible participation in the centre's  spin polarization process. Clearly, the model of the Jahn-Teller effect in the NV$^-$ centre needs to be developed in greater detail to fully understand the temperature dependence of the centre's properties and the role of the Jahn-Teller effect in spin polarization.

\section{Conclusion}

In this article, the molecular model of the NV$^-$ centre's electronic structure was developed in full. Through the explicit analysis of the Coulomb repulsion interaction, an insight was gained into the many competing factors that influence the energetic ordering of the singlet states of the ground and first excited MO configurations. Using a proof of the positive definite nature of the Coulomb repulsion separation of the states of the ground MO configuration, the simplest consistent orbital structure was adopted and the avenues of further investigation and routes to confirmation identified. Having obtained the orbital structure, spin-orbit and spin-spin interactions were treated using perturbation theory and the centre's fine structure and first order state couplings were determined. The results were expressed in terms of the simplest set of one- and two-electron reduced density matrix elements, allowing direct insight into which parameters were known and how to pursue the remaining unknown parameters. The calculation of the matrix representations of the centre's interactions with electric, magnetic and strain fields allowed the efficient discussion of the effects of strain on the centre's fine structure. The matrix representations also formed the basis of a short discussion of the temperature dependence of the centre's fine structure and motivated the central role that the dynamic Jahn-Teller effect plays in the centre's dynamics.

\ack
This work was supported by the Australian Research Council under the
Discovery Project scheme DP0986635.

\appendix
\section*{Appendix}
\setcounter{section}{1}

\Table{\label{tab:holes} The configuration states expressed in terms of holes rather than electrons. Note that the hole formulism has not been used in this article and these expressions are provided to assist the readers who have previously used the hole formulism.}
\br
 & & $\Phi_{j,k;S,m_s}^c$ & & & $\Phi_{n,j,k}^{so}$ & & \\
\mr
$e^2$ & $^3A_2$ & $\Phi_{A_2;1,0}^c$ & = & $\frac{1}{\sqrt{2}}(\ket{e_x\bar{e}_y}+\ket{\bar{e}_xe_y})$ & $\Phi_{1,A_1}^{so}$ & = & $\Phi_{A_2;1,0}^c$ \\
  & & $\Phi_{A_2;1,1}^c$ & = & $\ket{\bar{e}_x\bar{e}_y}$ & $\Phi_{2,E,x}^{so}$ & = & $\frac{-1}{\sqrt{2}}(-\Phi_{A_2;1,1}^c+\Phi_{A_2;1,-1}^c)$ \\
  & & $\Phi_{A_2;1,-1}^c$ & = & $\ket{e_xe_y}$ & $\Phi_{2,E,y}^{so}$ & = & $\frac{-i}{\sqrt{2}}(\Phi_{A_2;1,1}^c+\Phi_{A_2;1,-1}^c)$ \\
 & $^1E$ & $\Phi_{E,x;0,0}^c$ & = & $\frac{1}{\sqrt{2}}(\ket{e_y\bar{e}_y}-\ket{e_x\bar{e}_x})$ & $\Phi_{3,E,x}^{so}$ & = & $\Phi_{E,x;0,0}^{c}$ \\
  & & $\Phi_{E,y;0,0}^c$ & = & $\frac{1}{\sqrt{2}}(\ket{e_x\bar{e}_y}-\ket{\bar{e}_xe_y})$ & $\Phi_{3,E,y}^{so}$ & = & $\Phi_{E,y;0,0}^{c}$ \\
  & $^1A_1$ & $\Phi_{A_1;0,0}^c$ & = & $\frac{1}{\sqrt{2}}(\ket{e_y\bar{e}_y}+\ket{e_x\bar{e}_x})$ & $\Phi_{4,A_1}^{so}$ & = & $\Phi_{A_1;0,0}^c$ \\
 $a_1e$ & $^3E$ & $\Phi_{E,x;1,0}^c$ & = & $\frac{1}{\sqrt{2}}(\ket{\bar{a}_1e_x}+\ket{a_1\bar{e}_x})$ & $\Phi_{5,E,x}^{so}$ & = & $\frac{1}{2}\left[-i(\Phi_{E,x;1,1}^c+\Phi_{E,x;1,-1}^c))\right.$ \\
 & & & & & & & $\left.-(-\Phi_{E,y;1,1}^c+\Phi_{E,y;1,-1}^c)\right]$ \\
 & & $\Phi_{E,y;1,0}^c$ & = & $\frac{1}{\sqrt{2}}(\ket{\bar{a}_1e_y}+\ket{a_1\bar{e}_y})$ & $\Phi_{5,E,y}^{so}$ & = & $\frac{1}{2}\left[-(-\Phi_{E,x;1,1}^c+\Phi_{E,x;1,-1}^c)\right.$ \\
 & & & & & & & $\left.+i(\Phi_{E,y;1,1}^c+\Phi_{E,y;1,-1}^c)\right]$ \\
 & & $\Phi_{E,x;1,1}^c$ & = & $\ket{\bar{a}_1\bar{e}_x}$ & $\Phi_{6,E,x}^{so}$ & = & $-\Phi_{E,y;1,0}^c$ \\
 & & $\Phi_{E,y;1,1}^c$ & = & $\ket{\bar{a}_1\bar{e}_y}$ & $\Phi_{6,E,y}^{so}$ & = & $\Phi_{E,x;1,0}^c$ \\
 & & $\Phi_{E,x;1,-1}^c$ & = & $\ket{a_1e_x}$ & $\Phi_{7,A_2}^{so}$ & = & $\frac{1}{2}\left[(-\Phi_{E,x;1,1}^c+\Phi_{E,x;1,-1}^c)\right.$ \\
 & & & & & & & $\left.+i(\Phi_{E,y;1,1}^c+\Phi_{E,y;1,-1}^c)\right]$ \\
 & & $\Phi_{E,y;1,-1}^c$ & = & $\ket{a_1e_y}$ & $\Phi_{8,A_1}^{so}$ & = & $\frac{1}{2}\left[-i(\Phi_{E,x;1,1}^c+\Phi_{E,x;1,-1}^c))\right.$ \\
 & & & & & & & $\left.+(-\Phi_{E,y;1,1}^c+\Phi_{E,y;1,-1}^c)\right]$ \\
 & $^1E'$ & $\Phi_{E',x;0,0}^c$ & = & $\frac{1}{\sqrt{2}}(\ket{\bar{a}_1e_x}-\ket{a_1\bar{e}_x})$ & $\Phi_{9,E,x}^{so}$ & = & $\Phi_{E',x;0,0}^{c}$ \\
 & & $\Phi_{E',y;0,0}^c$ & = & $\frac{1}{\sqrt{2}}(\ket{\bar{a}_1e_y}-\ket{a_1\bar{e}_y})$ & $\Phi_{9,E,y}^{so}$ & = & $\Phi_{E',y;0,0}^{c}$ \\
\br
\endTable

\begin{table}
\caption{\label{tab:couplingcoef} The electronic coupling coefficients correct to first order in spin-orbit and spin-spin. Parameters are as defined in \tref{tab:parameters}.}
\begin{indented}
\item[]\begin{tabular}{lclll}
\br
$s_{n,m}$ & & $s_{n,m}^{(0)}$ & $s_{n,m}^{(1)}$ & $s_{n,m}^{(2)}$ \\
\mr
$s_{1,1}$ & = & 1 & - & - \\
$s_{1,4}$ & = & - & $-2i\frac{\lambda_\parallel}{E_{A_1;0}}$ & -   \\
$s_{1,8}$ & = & - & $-\sqrt{2}\frac{\lambda_{\perp}+D_{1,E,2}}{E_{E;1}}$ & -  \\
$s_{2,2}$ & = & 1 & - & - \\
$s_{2,3}$ & = & - & - & $iN_\kappa\kappa\frac{\lambda_\perp}{E_{E;0}}$   \\
$s_{2,5}$ & = & - & $-\sqrt{2}N_\eta\frac{D_{1,E,1}}{E_{E;1}}$ & $-N_\eta\eta\frac{\lambda_\perp-D_{1,E,2}}{D_{E;1}}$   \\
$s_{2,6}$ & = & - & $ N_\eta\frac{\lambda_\perp-D_{1,E,2}}{E_{E;1}}$ & $-\sqrt{2}N_\eta\eta\frac{D_{1,E,1}}{E_{E;1}}$  \\
$s_{2,9}$ & = & - & $-iN_\kappa\frac{\lambda_\perp}{E_{E';0}}$ & -  \\
$s_{3,3}$ & = & $N_\kappa$ & - & - \\
$s_{3,2}$ & = & - & - & $iN_\kappa\kappa\frac{\lambda_\perp}{E_{E;0}}$  \\
$s_{3,5}$ & = & - & $-i\sqrt{2}N_\kappa N_\eta\frac{\lambda_\perp}{E_{E;1}-E_{E;0}}$ & -  \\
$s_{3,6}$ & = & - & - & $-iN_\kappa N_\eta\frac{\kappa\lambda_\parallel+\sqrt{2}\eta\lambda_\perp}{E_{E:1}-E_{E:0}}$  \\
$s_{3,9}$ & = & - & $-N_\kappa\kappa$ & - \\
$s_{4,4}$ & = & 1 & - & - \\
$s_{4,1}$ & = & - & $-2i\frac{\lambda_\parallel}{E_{A_1;0}}$ & - \\
$s_{4,8}$ & = & - & $i\sqrt{2}\frac{\lambda_\perp}{E_{E;1}-E_{A_1;0}}$ & - \\
$s_{5,5}$ & = & $N_\eta$ & - & - \\
$s_{5,2}$ & = & - & $\sqrt{2}N_\eta\frac{D_{1,E,1}}{E_{E;1}}$ & $N_\eta\eta\frac{\lambda_\perp-D_{1,E,2}}{E_{E:1}}$ \\
$s_{5,3}$ & = & - & $-i\sqrt{2}N_\kappa N_\eta\frac{\lambda_\perp}{E_{E:1}-E_{E:0}}$ & -  \\
$s_{5,6}$ & = & - & $-N_\eta\eta$ & -  \\
$s_{5,9}$ & = & - & - & $iN_\kappa N_\eta\frac{\eta\lambda_\parallel+\sqrt{2}\kappa\lambda_\perp}{E_{E';0}-E_{E;1}}$ \\
$s_{6,6}$ & = & $N_\eta$ & - & - \\
$s_{6,2}$ & = & - & $-N_\eta\frac{\lambda_\perp-D_{1,E,2}}{E_{E:1}}$ & $\sqrt{2}N_\eta\eta\frac{D_{1,E,1}}{E_{E;1}}$  \\
$s_{6,3}$ & = & - & - &  $-iN_\kappa N_\eta\frac{\kappa\lambda_\parallel+\sqrt{2}\eta\lambda_\perp}{E_{E;1}-E_{E;0}}$ \\
$s_{6,5}$ & = & - & $N_\eta\eta$ & -  \\
$s_{6,9}$ & = & - & $-iN_\kappa N_\eta\frac{\lambda_\parallel}{E_{E';0}-E_{E:1}}$ & - \\
$s_{7,7}$ & = & 1 & - & - \\
$s_{8,8}$ & = & 1 & - & - \\
$s_{8,1}$ & = & - & $\sqrt{2}\frac{\lambda_{\perp}+D_{1,E,2}}{E_{E;1}}$ & -  \\
$s_{8,4}$ & = & - & $i\sqrt{2}\frac{\lambda_\perp}{E_{E;1}-E_{A_1;0}}$ & -  \\
$s_{9,9}$ & = & $N_\kappa$ & - & - \\
$s_{9,2}$ & = & - & $-iN_\kappa\frac{\lambda_\perp}{E_{E';0}}$ & -  \\
$s_{9,3}$ & = & - & $N_\kappa\kappa$ & - \\
$s_{9,5}$ & = & - & - & $iN_\kappa N_\eta\frac{\eta\lambda_\parallel+\sqrt{2}\kappa\lambda_\perp}{E_{E';0}-E_{E;1}}$  \\
$s_{9,6}$ & = & - & $-iN_\kappa N_\eta\frac{\lambda_\parallel}{E_{E';0}-E_{E;1}}$ & -  \\
\br
\end{tabular}
\end{indented}
\end{table}

\fulltable{\label{tab:orbitaloperators1}Combined matrix representation of $\hat{O}_{A_1}$ and $\hat{O}_{A_2}$ orbital operators  in the basis $\{\Phi_{1,A_1}^{so},\Phi_{2,E,x}^{so},\Phi_{2,E,y}^{so},\Phi_{3,E,x}^{so},\Phi_{3,E,y}^{so},\Phi_{4,A_1}^{so},\Phi_{5,E,x}^{so},\Phi_{5,E,y}^{so},\Phi_{6,E,x}^{so},\Phi_{6,E,y}^{so},
$ $\Phi_{7,A_2}^{so},\Phi_{8,A_1}^{so},\Phi_{9,E,x}^{so},\Phi_{9,E,y}^{so}\}$. To obtain the matrix representation of $\hat{O}_{A_1}$, set $O_{A_2} \rightarrow 0$,  $O_{a,A_1} \rightarrow 2(\rdm{a_1}{V_{A_1}}{a_1}+\rdm{e}{V_{A_1}}{e})$, and $O_{b,A_1} \rightarrow \rdm{a_1}{V_{A_1}}{a_1}+3\rdm{e}{V_{A_1}}{e}$. To obtain the matrix representation of $\hat{O}_{A_2}$, set $O_{A_2} \rightarrow \rdm{e}{V_{A_2}}{e}$, $O_{a,A_1} \rightarrow  0$, and  $O_{b,A_1} \rightarrow  0$.}
\br
 $O_{a,A_1}$  & 0 & 0 & 0 & 0 & 0 & 0 & 0 & 0 & 0 & 0 & 0 & 0 & 0 \\
 0 & $O_{a,A_1}$  & 0 & 0 & 0 & 0 & 0 & 0 & 0 & 0 & 0 & 0 & 0 & 0 \\
 0 & 0 & $O_{a,A_1}$  & 0 & 0 & 0 & 0 & 0 & 0 & 0 & 0 & 0 & 0 & 0 \\
 0 & 0 & 0 & $O_{a,A_1}$  & $2O_{A_2}$ & 0 & 0 & 0 & 0 & 0 & 0 & 0 & 0 & 0 \\
 0 & 0 & 0 & $-2O_{A_2}$ & $O_{a,A_1}$  & 0 & 0 & 0 & 0 & 0 & 0 & 0 & 0 & 0 \\
 0 & 0 & 0 & 0 & 0 & $O_{a,A_1}$  & 0 & 0 & 0 & 0 & 0 & 0 & 0 & 0 \\
 0 & 0 & 0 & 0 & 0 & 0 & $O_{b,A_1}$  & $O_{A_2}$ & 0 & 0 & 0 & 0 & 0 & 0 \\
 0 & 0 & 0 & 0 & 0 & 0 & $-O_{A_2}$ & $O_{b,A_1}$  & 0 & 0 & 0 & 0 & 0 & 0 \\
 0 & 0 & 0 & 0 & 0 & 0 & 0 & 0 & $O_{b,A_1}$  & $-O_{A_2}$ & 0 & 0 & 0 & 0 \\
 0 & 0 & 0 & 0 & 0 & 0 & 0 & 0 & $O_{A_2}$ & $O_{b,A_1}$  & 0 & 0 & 0 & 0 \\
 0 & 0 & 0 & 0 & 0 & 0 & 0 & 0 & 0 & 0 & $O_{b,A_1}$  & $-O_{A_2}$ & 0 & 0 \\
 0 & 0 & 0 & 0 & 0 & 0 & 0 & 0 & 0 & 0 & $O_{A_2}$ & $O_{b,A_1}$ & 0 & 0 \\
 0 & 0 & 0 & 0 & 0 & 0 & 0 & 0 & 0 & 0 & 0 & 0 & $O_{b,A_1}$  & $-O_{A_2}$ \\
 0 & 0 & 0 & 0 & 0 & 0 & 0 & 0 & 0 & 0 & 0 & 0 & $O_{A_2}$ & $O_{b,A_1}$ \\
\br
\endfulltable

\fulltable{\label{tab:orbitaloperators2}Combined matrix representation of $\hat{O}_{E,x}$ and $\hat{O}_{E,y}$ orbital operators  in the basis $\{\Phi_{1,A_1}^{so},\Phi_{2,E,x}^{so},\Phi_{2,E,y}^{so},\Phi_{3,E,x}^{so},\Phi_{3,E,y}^{so},\Phi_{4,A_1}^{so},\Phi_{5,E,x}^{so},\Phi_{5,E,y}^{so},\Phi_{6,E,x}^{so},\Phi_{6,E,y}^{so},
$ $\Phi_{7,A_2}^{so},\Phi_{8,A_1}^{so},\Phi_{9,E,x}^{so},\Phi_{9,E,y}^{so}\}$. To obtain the matrix representation of $\hat{O}_{E,x}$, set $O_{a,x} \rightarrow \frac{1}{\sqrt{2}}\rdm{a_1}{V_{E}}{e}$, $O_{a,y} \rightarrow 0$, $O_{b,x} \rightarrow \frac{1}{\sqrt{2}}\rdm{e}{V_{E}}{e}$, and $ O_{b,y} \rightarrow 0$. To obtain the matrix representation of $\hat{O}_{E,y}$, set $O_{a,y} \rightarrow \frac{1}{\sqrt{2}}\rdm{a_1}{V_{E}}{e}$, $O_{a,x} \rightarrow 0$, $O_{b,y} \rightarrow \frac{1}{\sqrt{2}}\rdm{e}{V_{E}}{e}$, and $ O_{b,x} \rightarrow 0$.}
\br
 0 & 0 & 0 & 0 & 0 & 0 & 0 & 0 & $O_{a,x}$ & $O_{a,y}$ & 0 & 0 & 0 & 0 \\
 0 & 0 & 0 & 0 & 0 & 0 & $-\frac{O_{a,x}}{\sqrt{2}}$ & $\frac{O_{a,y}}{\sqrt{2}}$ & 0 & 0 & $-\frac{O_{a,y}}{\sqrt{2}}$ & $\frac{O_{a,x}}{\sqrt{2}}$ & 0 & 0 \\
 0 & 0 & 0 & 0 & 0 & 0 & $\frac{O_{a,y}}{\sqrt{2}}$ & $\frac{O_{a,x}}{\sqrt{2}}$ & 0 & 0 & $\frac{O_{a,x}}{\sqrt{2}}$ & $\frac{O_{a,y}}{\sqrt{2}}$ & 0 & 0 \\
 0 & 0 & 0 & 0 & 0 & $2 O_{b,x}$ & 0 & 0 & 0 & 0 & 0 & 0 & $-O_{a,x}$ & $O_{a,y}$ \\
 0 & 0 & 0 & 0 & 0 & $2O_{b,y}$ & 0 & 0 & 0 & 0 & 0 & 0 & $O_{a,y}$ & $O_{a,x}$ \\
 0 & 0 & 0 & $2 O_{b,x}$ & $2O_{b,y}$ & 0 & 0 & 0 & 0 & 0 & 0 & 0 & $O_{a,x}$ & $O_{a,y}$ \\
 0 & $-\frac{O_{a,x}^\ast}{\sqrt{2}}$ & $\frac{O_{a,y}^\ast}{\sqrt{2}}$ & 0 & 0 & 0 & 0 & 0 & 0 & 0 & $-O_{b,y}$ & $-O_{b,x}$ & 0 & 0 \\
 0 & $\frac{O_{a,y}^\ast}{\sqrt{2}}$ & $\frac{O_{a,x}^\ast}{\sqrt{2}}$ & 0 & 0 & 0 & 0 & 0 & 0 & 0 & $O_{b,x}$ & $-O_{b,y}$ & 0 & 0 \\
 $O_{a,x}^\ast$ & 0 & 0 & 0 & 0 & 0 & 0 & 0 & $O_{b,x}$ & $-O_{b,y}$ & 0 & 0 & 0 & 0 \\
 $O_{a,y}^\ast$ & 0 & 0 & 0 & 0 & 0 & 0 & 0 & $-O_{b,y}$ & $-O_{b,x}$ & 0 & 0 & 0 & 0 \\
 0 & $-\frac{O_{a,y}^\ast}{\sqrt{2}}$ & $\frac{O_{a,x}^\ast}{\sqrt{2}}$ & 0 & 0 & 0 & $-O_{b,y}$ & $O_{b,x}$ & 0 & 0 & 0 & 0 & 0 & 0 \\
 0 & $\frac{O_{a,x}^\ast}{\sqrt{2}}$ & $\frac{O_{a,y}^\ast}{\sqrt{2}}$ & 0 & 0 & 0 & $-O_{b,x}$ & $-O_{b,y}$ & 0 & 0 & 0 & 0 & 0 & 0 \\
 0 & 0 & 0 & $-O_{a,x}^\ast$ & $O_{a,y}^\ast$ & $O_{a,x}^\ast$ & 0 & 0 & 0 & 0 & 0 & 0 & $-O_{b,x}$ & $O_{b,y}$ \\
 0 & 0 & 0 & $O_{a,y}^\ast$ & $O_{a,x}^\ast$ & $O_{a,y}^\ast$ & 0 & 0 & 0 & 0 & 0 & 0 & $O_{b,y}$ & $O_{b,x}$ \\
\br
\endfulltable

\Table{\label{tab:spinoperators}Combined matrix representation of the components of the total spin operator $\vec{S}$  in the basis $\{\Phi_{1,A_1}^{so},\Phi_{2,E,x}^{so},\Phi_{2,E,y}^{so},\Phi_{3,E,x}^{so},\Phi_{3,E,y}^{so},\Phi_{4,A_1}^{so},\Phi_{5,E,x}^{so},\Phi_{5,E,y}^{so},\Phi_{6,E,x}^{so},\Phi_{6,E,y}^{so},
\Phi_{7,A_2}^{so},\Phi_{8,A_1}^{so},$ $\Phi_{9,E,x}^{so},\Phi_{9,E,y}^{so}\}$. To obtain the matrix representation of $\hat{S}_x$, set $S_x \rightarrow \hbar$ and $S_y$, $S_z\rightarrow 0$. To obtain the matrix representation of $\hat{S}_y$, set $S_y \rightarrow \hbar$ and $S_x$, $S_z\rightarrow 0$. To obtain the matrix representation of $\hat{S}_z$, set $S_z \rightarrow \hbar$ and $S_x$, $S_y\rightarrow 0$.}
\br
 0 & $iS_y$ & $-iS_x$ & 0 & 0 & 0 & 0 & 0 & 0 & 0 & 0 & 0 & 0 & 0 \\
 $-iS_y$ & 0 & $-iS_z$ & 0 & 0 & 0 & 0 & 0 & 0 & 0 & 0 & 0 & 0 & 0 \\
 $iS_x$ & $iS_z$ & 0 & 0 & 0 & 0 & 0 & 0 & 0 & 0 & 0 & 0 & 0 & 0 \\
 0 & 0 & 0 & 0 & 0 & 0 & 0 & 0 & 0 & 0 & 0 & 0 & 0 & 0 \\
 0 & 0 & 0 & 0 & 0 & 0 & 0 & 0 & 0 & 0 & 0 & 0 & 0 & 0 \\
 0 & 0 & 0 & 0 & 0 & 0 & 0 & 0 & 0 & 0 & 0 & 0 & 0 & 0 \\
 0 & 0 & 0 & 0 & 0 & 0 & 0 & $iS_z$ & $\frac{i}{\sqrt{2}}S_y$ & $\frac{i}{\sqrt{2}}S_x$ & 0 & 0 & 0 & 0 \\
 0 & 0 & 0 & 0 & 0 & 0 & $-iS_z$ & 0 & $\frac{i}{\sqrt{2}}S_x$ & $\frac{-i}{\sqrt{2}}S_y$ & 0 & 0 & 0 & 0 \\
 0 & 0 & 0 & 0 & 0 & 0 & $\frac{-i}{\sqrt{2}}S_y$ & $\frac{-i}{\sqrt{2}}S_x$ & 0 & 0 & $\frac{-i}{\sqrt{2}}S_x$ & $\frac{i}{\sqrt{2}}S_y$ & 0 & 0 \\
 0 & 0 & 0 & 0 & 0 & 0 & $\frac{-i}{\sqrt{2}}S_x$ & $\frac{-i}{\sqrt{2}}S_y$ & 0 & 0 & $\frac{-i}{\sqrt{2}}S_y$ & $\frac{-i}{\sqrt{2}}S_x$ & 0 & 0 \\
 0 & 0 & 0 & 0 & 0 & 0 & 0 & 0 & $\frac{i}{\sqrt{2}}S_x$ & $\frac{i}{\sqrt{2}}S_y$ & 0 & $iS_z$ & 0 & 0 \\
 0 & 0 & 0 & 0 & 0 & 0 & 0 & 0 & $\frac{-i}{\sqrt{2}}S_y$ & $\frac{i}{\sqrt{2}}S_x$ & $-iS_z$ & 0 & 0 & 0 \\
 0 & 0 & 0 & 0 & 0 & 0 & 0 & 0 & 0 & 0 & 0 & 0 & 0 & 0 \\
 0 & 0 & 0 & 0 & 0 & 0 & 0 & 0 & 0 & 0 & 0 & 0 & 0 & 0 \\
\br
\endTable

\section*{References}


\begin{thebibliography}{99}
\bibitem{qip} Wrachtrup J and Jelezko F 2006 {\it \JPCM} {\bf 18} S807
\bibitem{bio} Fu C-C, Lee H-Y, Chen K, Lim T-S, Wu H-Y, Lin P-K, Wei P-K, Tsoa P-H, Chang H C and Fann W 2007 {\it PNAS} {\bf 104} 727
\nonum Chang Y-R \etal 2008 {\it Nature Nanotech} {\bf 3} 284
\nonum Neugart F, Zappe A, Jelezko F, Tietz C, Boudou J P, Krueger A and Wrachtrup J 2007 {\it Nano Letters} {\bf 7} 3588
\bibitem{qkd} Beveratos A, Brouri R, Gacoin T, Villing A, Poizat J P and Grangier P 2002 {\it \PRL} {\bf 89} 187901
\bibitem{comp} Jelezko F, Gaebel T, Popa I, Domhan M, Gruber A and Wrachtrup J 2004 {\it \PRL} {\bf  93} 130501
\nonum Gurudev Dutt M V, Childress L, Jiang L, Togan E, Maze J, Jelezko F, Zibrov A S, Hemmer P R and Lukin M D 2007 {\it Science} {\bf 316} 1312
\nonum Ladd T D, Jelezko F, Laflamme R, Nakamura Y, Monroe C and O'Brien J L 2010 {\it Nature} {\bf 464} 45
\bibitem{qswitch} Greentree A D, Salzman J, Prawer S and Hollenberg L C L 2006 {\it \PR} A {\bf 73} 013818
\bibitem{mag} Balasubramanian G \etal 2008 {\it Nature} {\bf 455} 648
\nonum Maze J R \etal 2008 {\it Nature} {\bf 455} 644
\nonum Degen C L 2008 {\it Appl. Phys. Lett.} {\bf 92} 243111
\nonum Steinert S, Dolde F, Neumann P, Aird A, Naydenov B, Balasubramanian G, Jelezko F and Wrachtrup J 2010 {\it Review of Scientific Instruments} {\bf 81} 043705
\bibitem{efield} Dolde F \etal 2010 {\it In Preparation}
\bibitem{decoherence} Cole J H and Hollenberg L C L 2009 {\it Nanotechnology} {\bf 20} 495401
\nonum Hall L T, Cole J H, Hall C D and Hollenberg L C L 2009 {\it \PRL} {\bf 103} 220802
\bibitem{singlephoton} Kurtsiefer C, Mayer S, Zarda P and Weinfurter H 2000 {\it \PRL} {\bf 85} 290
\bibitem{coherence} Balasubramanian G \etal 2009 {\it Nature Materials} {\bf 8} 383
\bibitem{coupling} Neumann P, Mizouchi N, Rempp F, Hemmer P, Watanabe H, Yamasaki S, Jacques V, Gaebel T, Jelezko F and Wrachtrup J 2008 {\it Science} {\bf 320} 1326
\bibitem{readout} Harrison J, Sellars M J and Manson N B 2004 {\it J. Lumin.} {\bf 107} 245
\nonum Jelezko F and Wrachtrup J 2004 {\it \JPCM} {\bf 16} 1089
\bibitem{dupreez} du Preez L 1965 {\it PhD thesis} University of Witwatersand
\bibitem{infrared} Rogers L J, Armstrong S, Sellars M J and Manson N B 2008 {\it \NJP} {\bf 10} 103024
\bibitem{ground} Reddy N R S, Manson N B and Krausz E R 1987 {\it J. Lumin.} {\bf 38} 46
\bibitem{excitedstatestrain} Batalov A, Jacques V, Kaiser F, Siyushev P, Neumann P, Rogers L J, McMurtrie R L, Manson N B, Jelezko F and Wrachtrup J 2009 {\it \PRL} {\bf 102} 195506
\bibitem{excitedstatezeeman} Neumann P \etal 2009 {\it \NJP} {\bf 11} 013017
\bibitem{tamarat} Tamarat P H \etal 2006 {\it \NJP} {\bf 10} 045004
\bibitem{averaging} Rogers L J \etal 2009 {\it \NJP} {\bf 11} 063007
\bibitem{jahnteller} Fu K M C, Santori C, Barclay P E, Rogers L J, Manson N B and Beausoleil R G 2009 {\it \PRL} {\bf 103} 256404
\bibitem{loubser} Loubser J H N and van Wyk J A 1978 {\it Rep. Prog. Phys.} {\bf 41} 1203
\bibitem{xing} He X-F, Manson N B and Fisk P T H 1993 {\it \PR} B {\bf 47} 8816
\bibitem{lenef} Lenef A and Rand S C 1996 {\it \PR} B {\bf 53} 13441
\bibitem{manson} Manson N B, Harrison J P and Sellars M J 2006 {\it \PR} B {\bf 74} 104304
\bibitem{vanoort} van Oort E and Glasbeek M 1990 {\it Chem. Phys. Lett.} {\bf 168} 529
\bibitem{gfactor} Felton S, Edmonds A M, Newton M E, Martineau P M, Fisher D, Twitchen D J and Baker J M 2009 {\it \PR} B {\bf 79} 075203
\bibitem{ab initio} Goss J P, Jones R, Breyer S J, Briddon P R and Oberg S 1996 {\it \PRL} {\bf 77} 3041
\nonum Luszczek M, Laskowski R and Horodecki P 2004 {\it Physica} B {\bf 348} 292
\nonum Larsson J A and Delaney P 2008 {\it \PR} B \textbf{77} 165201
\nonum Lin C, Wang Y, Chang H, Hayashi M and Lin S H 2008 {\it \JCP} {\bf 129} 124714
\nonum Gali A, Janzen E, Deak P, Kresse G and Kaxiras E 2009 {\it \PRL} {\bf 103} 186404
\bibitem{gali2} Gali A, Fyta M and Kaxiras E 2008 {\it \PR} B {\bf 77} 155206
\bibitem{prl}  Hossain F M, Doherty M W, Wilson H and Hollenberg L C L 2008 {\it \PRL} {\bf 101} 226403
\bibitem{delaney} Delaney P, Greer J C and Larsson J A 2010 {\it Nano Letters} {\bf 10} 610
\bibitem{gali3} Ma Y, Rohlfing M and Gali A 2010 {\it \PR} B {\bf 81} 041204
\bibitem{cornwell} Cornwell J F 1997 {\it Group Theory in Physics: an introduction} (London: Academic Press Inc.)
\bibitem{proof} Jackson J D 1998 {\it Classical Electrodynamics} (New York: Wiley)
\bibitem{maze} Rios J M 2010 {\it PhD thesis} Havard University
\bibitem{paul} Delaney P and Larrson J A 2010 {\it In Preparation}
\bibitem{singletlevel} Drabenstedt A, Fleury L, Tietz C, Jelezko F, Kilin S, Nizovtzev A and Wrachtrup J 1999 {\it \PR} B {\bf 60} 11503
\bibitem{tinkham} Tinkham M 2003 {\it Group Theory and Quantum Mechanics} (New York: Dover Publications).
\bibitem{vibcoupling} Fischer G 1984 {\it Vibronic Coupling: The Interaction between the Electronic and Nuclear Motions} (London: Academic Press Inc.)
\bibitem{stoneham} Stoneham A M 1975 {\it Theory of Defects in Solids} (Oxford: Oxford University Press)


\end{thebibliography}
\end{document}